\documentclass[11pt]{article}
\pdfoutput=1
\usepackage{jheppub}
\usepackage{amsfonts}
\usepackage{amscd}
\usepackage{amssymb}
\usepackage{amsmath,bbm}
\usepackage{graphicx}
\usepackage{epsfig}
\usepackage{latexsym}
\usepackage{mathtools}
\usepackage{hyperref}
\usepackage[vcentermath]{youngtab}
\hypersetup{
    colorlinks=true,
   linkcolor=black,
    citecolor=black,
    filecolor=black,
    urlcolor=black,}
    
\def \be  {\begin{equation}}
\def \ee  {\end{equation}}
\def \bea {\begin{equation}\begin{aligned}}
\def \eea {\end{aligned}\end{equation}}
\def \ba  {\begin{eqnarray}}
\def \ea  {\end{eqnarray}}
\def \bb  {\begin{thebibliography}}
\def \eb  {\end{thebibliography}}
\def \lab #1 {\label{#1}}

\newcommand\cF{\mathcal{F}}

\newcommand\cH{\mathcal{H}}
\newcommand\cI{\mathcal{I}}

\newcommand\cM{\mathcal{M}}
\newcommand\cN{\mathcal{N}}
\newcommand\cO{\mathcal{O}}

\newcommand\cS{\mathcal{S}}

\newcommand\cW{\mathcal{W}}

\newcommand\al{\alpha}

\newcommand\gf{\mathfrak{g}}
\newcommand\sutwo{\mathfrak{su}(2)}

\newcommand\bZ{\mathbb{Z}}

\newcommand\lb{\lambda}

\newcommand\la{\langle}
\newcommand\ra{\rangle}
\newcommand\del{\partial}

\newcommand\tr{\mathrm{Tr}}

\newcommand\ep{\epsilon}

\newcommand{\PE}[1]{ \mathrm{PE}\left[ #1 \right] }

\title{The Superconformal Index of the (2,0) Theory with Defects}
\author[a]{Mathew Bullimore\,}
\affiliation[a]{Institute for Advanced Study, Einstein Dr., Princeton, NJ 08540, USA}
\author[b]{Hee-Cheol Kim\,}
\affiliation[b]{Perimeter Institute for Theoretical Physics, Waterloo, Ontario N2L 2Y5, Canada}

\abstract{We compute the supersymmetric partition function of the six-dimensional $(2,0)$ theory of type $A_{N-1}$ on $S^1 \times S^5$ in the presence of both codimension two and codimension four defects. We concentrate on a limit of the partition function depending on a single parameter. From the allowed supersymmetric configurations of defects we find a precise match with the characters of irreducible modules of $W_N$ algebras and affine Lie algebras of type $A_{N-1}$.}

\begin{document}
\maketitle


\section{Introduction}\label{sec:introduction}

A remarkable prediction of string theory is the existence of interacting $\cN=(2,0)$ superconformal quantum field theories in six dimensions, which underpins many recent developments in mathematical physics. For example, compactifications on Riemann surfaces lead to a rich class $\cS$ of $\cN=2$ supersymmetric quantum field theories in four dimensions~\cite{Gaiotto:2009we,Gaiotto:2009hg}. A consequence of the six-dimensional perspective is that supersymmetric partition functions of these theories on the manifolds $S^4$ and $S^1 \times S^3$ are related to conformal field theories~\cite{Alday:2009aq} and topological field theories~\cite{Gadde:2009kb,Gadde:2011ik} respectively. Furthermore, compactifications on three-manifolds~\cite{Dimofte:2011ju,Dimofte:2011py} and four-manifolds~\cite{Gadde:2013sca} also lead to interesting classes of supersymmetric theories in lower dimensions and new connections between mathematics and physics.

In this paper, we will consider the supersymmetric partition function of the $\cN=(2,0)$ theory itself on $S^1 \times S^5$, which is closely related to the 6d superconformal index. Although this partition function cannot be computed directly in six dimensions, due to the absence of a useful lagrangian formulation, it has been computed recently using maximally supersymmetric Yang-Mills theory in five dimensions on $S^5$~\cite{Kim:2012ava,Lockhart:2012vp,Kim:2012qf}\footnote{See also~\cite{Kallen:2012cs,Hosomichi:2012ek,Kallen:2012va,Imamura:2012xg,Imamura:2012bm} for closely related works on $S^5$ partition functions.}. It relies on the conjecture that the non-perturbative physics of 5d SYM allows us to extract non-trivial dynamics of the 6d (2,0) theory with circle compactification~\cite{Douglas:2010iu,Lambert:2010iw}.
The most important part of the dictionary is that the five-dimensional gauge coupling $g^2$ is related to the radius $\beta$ of $S^1$ by 
\be
g^2 = 2 \pi \beta \, .
\ee
so that strong coupling corresponds to large radius. Although this five-dimensional theory is non-renormalizable, it is conjectured that by including non-perturbative contributions in five dimensions, one can capture all of the protected states contributing to the 6d superconformal index. This is remarkable given that the 5d partition function is computed as an instanton expansion in powers of $e^{-4\pi^2/\beta}$, while the 6d superconformal index is naturally an expansion at large radius in powers of $e^{-\beta}$ with integer coefficients.

In this paper, we will extend these results to compute the partition function of the $(2,0)$ theory on $S^1 \times S^5$, or superconformal index, in the presence of extended defect operators. One motivation for this is that extended defect operators play an indispensable role in engineering defects of various kinds in compactifications to lower dimensions. As one example, compactification on a Riemann surface with codimension 2 defects sitting at punctures leads to flavor symmetries in the resulting four dimensional $\cN=2$ theory~\cite{Gaiotto:2009we,Gaiotto:2009hg}.

We will concentrate solely on the six dimensional $(2,0)$ theory of type $\mathfrak{g}=A_{N-1}$, which arises on the worldvolume of $N$ coincident M5 branes. There are codimension 4 defects from intersecting M2 branes and codimension 2 defects from intersecting M5 branes. Our working assumption is that the defects have the following classification:
\begin{enumerate}
\item Codimension 4 defects are labelled by a dominant integral weight $\lambda$ of $\gf$.
\item Codimension 2 defects are labelled by homomorphisms $\rho: \sutwo  \to \gf$.
\item Codimension 4 defects coincident with a codimension 2 defect of type $\rho$ are labelled by a dominant integral weight of the stabilizer of $\mathrm{Im}(\rho) \subset \gf$.
\end{enumerate}
For our computations, we will assume that defects wrapping $S^1$ have an effective description in 5d $\cN=2$ SYM theory as:
\begin{enumerate}
\item A supersymmetric Wilson line in the irreducible representation of $SU(N)$ with highest weight $\lambda$.
\item A surface defect obtained by coupling to the three-dimensional $\cN=4$ theory $T_{\rho}(SU(N))$. Alternatively, a monodromy defect whose monodromy is labelled by $\rho$.
\item A supersymmetric Wilson line in an irreducible representation of the unbroken gauge group in the presence of the monodromy $\rho$.
\end{enumerate}

Let us now provide a some more details about the computation. The most general superconformal index or partition function on $S^1 \times S^5$ preserves a single supercharge and its conjugate. It depends on five parameters corresponding to combinations of bosonic charges that commute with the supercharge. In the language of five-dimensional gauge theory on $S^5$, these parameters can be understood as follows. Firstly, there is the radius $\beta$ of $S^1$ which is related to the five-dimensional gauge coupling $g^2$ by the formula $g^2 = 2\pi \beta$. Secondly, there are three squashing parameters $\vec \omega = (\omega_1 , \omega_2, \omega_3)$ for the geometry of $S^5$. Finally, there is a real mass parameter $\mu$ for the adjoint hypermultiplet inside the $\cN=2$ vectormultiplet. 

We will furthermore concentrate on a limit where we tune the real mass parameter as follows
\be
\mu \to \frac{1}{2}(\omega_1+\omega_2-\omega_3) \, .
\ee
The partition function preserves an additional supercharge in this background and leads to dramatic simplifications in the answer~\cite{Kim:2012ava,Kim:2012qf,Beem:2014kka}. In the absence of defects, the partition function has been shown to coincide with the character of the vacuum module of the $W_N$ - algebra with central charge $c = (N-1) + N(N^2-1) \frac{(\omega_1+\omega_1)^2}{\omega_1\omega_2}$. This result has been interpreted recently in the context of chiral algebras~\cite{Beem:2014kka}. For this reason, we will refer to this background as the `chiral algebra' limit.

\begin{figure}[tb]
\centering
\includegraphics[width=13cm]{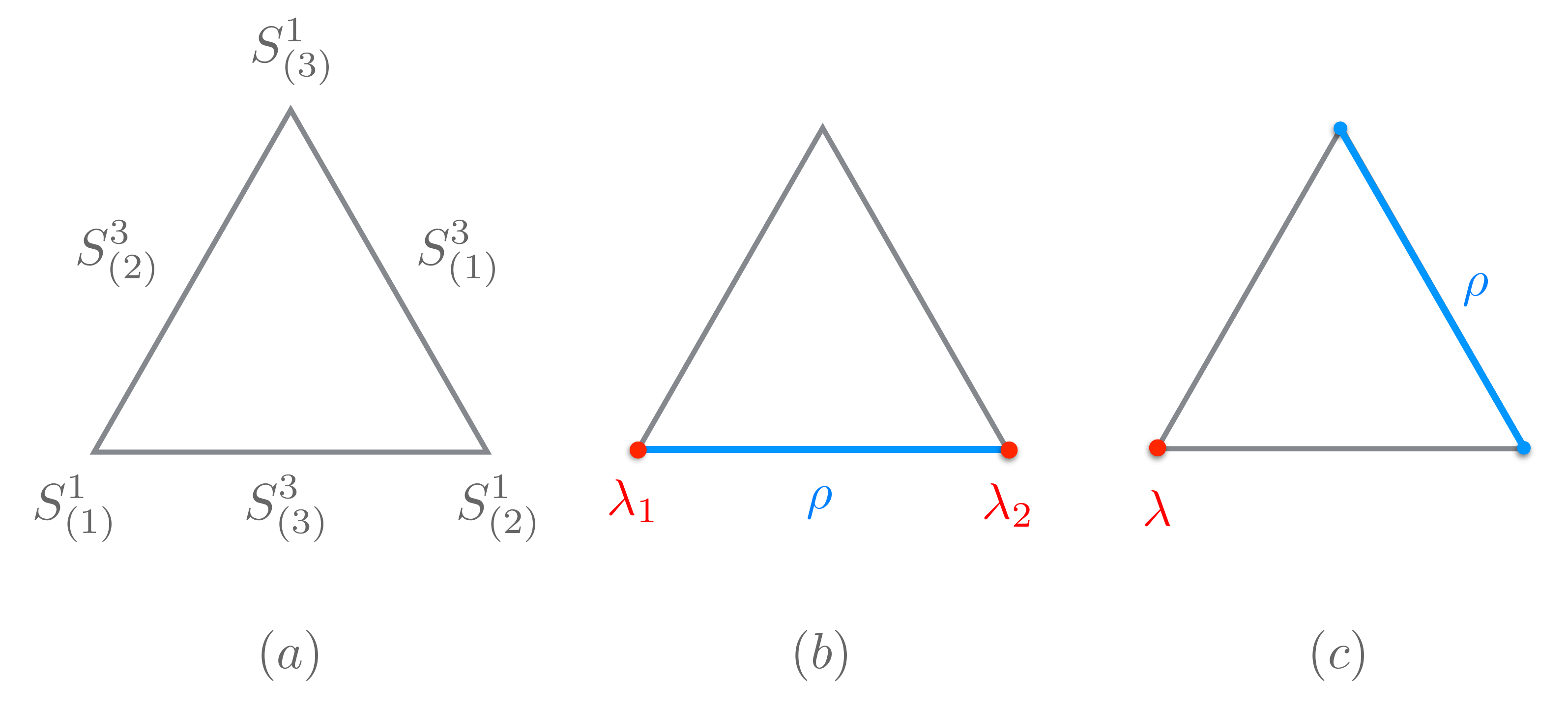}
\caption{\emph{A summary of the configurations of defects referred to in the introduction. Codimension 2 defects are shown in blue and codimension 4 defects are shown in red.}}
\label{fig:summary}
\end{figure}

To picture where we can add supersymmetric defects to the calculation it is convenient to picture $S^5$ as a $(S^1)^3$ fibration over a triangle - see figure~\ref{fig:summary}(a). This is described further in the main text. There are three distinguished circles $S^{1}_{(a)}$ that may support supersymmetric Wilson loops and three distinguished squashed spheres $S^3_{(a)}$ which can support supersymmetric surface defects. In the chiral algebra limit, the circle $S_{(3)}^1$ plays a distinguished role. We expect quantitatively different results depending on whether or not the supersymmetric defects wrap the particular circle $S^1_{(3)}$. The configurations preserving the supercharges of the chiral algebra limit are shown in figure~\ref{fig:summary}(b) and~\ref{fig:summary}(c). Let us discuss each case in turn.

The most general configurations of defects that are supported away from $S^1_{(3)}$ are shown in figure~\ref{fig:summary} (b). The summary of our results is as follows:
\begin{enumerate}
\item Adding supersymmetric Wilson loops in representations of highest weights $\lb_1$ and $\lb_2$ supported on $S^1_{(1)}$ and $S^1_{(2)}$ we find the characters of fully degenerate modules of the $W_N$ - algebra.
\item Adding a supersymmetric surface defect of type $\rho$ supported on $S^3_{(3)}$ we find the character of a semi-degenerate module of the $W_N$ - algebra. The non-degenerate modules correspond to the maximal case $\rho = [1^N]$.
\item Adding a supersymmetric surface defect of type $\rho$ together with supersymmetric Wilson loops in representations of the stabilizer of $\mathrm{Im}(\rho) \subset \mathfrak{g}$ with highest weights $\lb_1$ and $\lb_2$, we find further semi-degenerate modules with a more intricate structure of null states.
\end{enumerate}
This completely exhausts the spectrum of irreducible modules of the $W_N$ - algebra described in~\cite{Drukker:2010jp}. From perspective of chiral algebras, our computations provide evidence that combinations of supersymmetric defects orthogonal to the chiral algebra plane are realized by chiral vertex operators for the $W_N$ - algebra, as speculated in~\cite{Beem:2014kka}.

Let us now summarize what happens when a codimension two defect is supported on either $S^3_{(1)}$ or $S^3_{(2)}$ and hence wraps the distinguished circle $S^1_{(3)}$. We will present the result for $S^3_{(1)}$ as shown in figure~\ref{fig:summary}(c). The result for $S^{3}_{(2)}$ is obtained by simply interchanging $\omega_1 \leftrightarrow \omega_2$. Furthermore, we focus on supersymmetric surface defects labelled by the partition $\rho = [1^N]$. In this case we find
\begin{enumerate}
\item For a supersymmetric surface defect supported on $S^3_{(1)}$ we find the character of the  vacuum module of affine $\widehat{\mathfrak{su}}(N)$ at level $k = -N - \omega_1 /\omega_2$.
\item Adding a supersymmetric Wilson loop supported on $S^1_{(1)}$ in a representation of highest weight $\lambda$ we find the character of an irreducible module of the above with highest affine weight $\hat\lb = k \, \hat\omega_0 + \lb$.
\end{enumerate}
This is a small generalization of the conjecture for the chiral algebra associated to a codimension 2 defect in~\cite{Beem:2014kka}. For supersymmetric surface defects of generic type $\rho$, we would expect to find characters of modules of $W(\rho)$ - algebras, which are obtained from the affine algebra by Drinfeld-Sokolov reduction. From this point of view the $W_N$ - algebra is the special case $\rho = [N]$ corresponding to the absence of a defect. However, we could not evaluate the matrix integrals arising from localization in the generic case.

In summary, it is remarkable that the combinatorics and characters of irreducible modules of a large class of chiral algebras are in 1-1 correspondence with supersymmetric configurations of M2 and M5 branes on $S^1 \times S^5$.

We now summarize the contents of the paper. In section~\ref{sec:method} we explain how to compute the general superconformal index in the presence of defects using localization on $S^5$ and perform some computations relevant for the chiral algebra limit. In section~\ref{sec:walgebra} we evaluate the partition functions with configurations of defects relevant for $W_N$ - algebras, as in figure~\ref{fig:summary}(b). In section~\ref{sec:affine} we evaluate the partition functions with configurations of defects relevant for affine algebras, as shown in figure~\ref{fig:summary}(c). We conclude in section~\ref{sec:disc} with a discussion of some interesting directions for further study.


\section{Computational Method}\label{sec:method}

In this section, we fix our notation for the paper and explain the method for computing the superconformal index of interacting $(2,0)$ theories in the presence of codimension 2 and 4 defects. The reader interested only in the final results and the connections to characters of vertex operator modules of chiral algebras can safely turn to sections~\ref{sec:walgebra} and~\ref{sec:affine}.


\subsection{The Superconformal Index}\label{subsec:index}

Let us define our conventions for the six-dimensional $(2,0)$ superconformal algebra $\mathfrak{osp}(8^*|4)$. The maximal bosonic subalgebra is $\mathfrak{so}(2,6) \oplus \mathfrak{usp}(4)$ and we denote the corresponding Cartan generators by $(E,h_1,h_2,h_3,R_1,R_2)$. In particular, the generator $E$ corresponds to dilatations, $(h_1,h_2,h_3)$ to rotations in three orthogonal planes of $\mathbb{R}^6$ and $(R_1,R_2)$ to $R$-symmetry generators or equivalently rotations in two orthogonal two-planes of the transverse $\mathbb{R}^5$. 

In addition, there are supersymmetry generators $Q_{h_1,h_2,h_3}^{R_2,R_2}$ labelled by their charges under the bosonic subalgebra. The indices may take the values $\pm \frac{1}{2}$ but for brevity we denote these by $\pm$ in what follows. There are sixteen Poincar\'e supercharges with $h_1h_2h_3<0$ and sixteen conformal supercharges with $h_1h_2h_3>0$. In radial quantization, conjugation reverses $h_1,h_2,h_3,R_1,R_2$ and hence interchanges Poincar\'e and conformal supercharges.

The superconformal index can be defined as a trace over the Hilbert space of the theory in radial quantization~\cite{Kinney:2005ej}. For 6d SCFTs, it was first introduced in~\cite{Bhattacharya:2008zy}. Here, we define the superconformal index using the supercharge $Q \equiv Q_{---}^{++}$. Although all choices lead to an equivalent superconformal index, this choice has the feature that it is symmetric in the generators $h_1,h_2,h_3$. 
The superalgebra generated by this supercharge is 
\be
\{ Q,Q^\dagger \} = E - 2(R_1+R_2) - (h_1+h_2+h_3) \ ,
\ee
with the conjugate supercharge $Q^\dagger \equiv Q_{+++}^{--}$.
The superconformal index counts states in short representations annihilated by $Q$ and $Q^\dagger$ and therefore saturating the bound
\be
E \ge 2(R_1+R_2) + h_1+h_2+h_3 \ .
\label{bound}
\ee
The superconformal index is then defined by
\be
\cI = \tr_{H_Q} \left[ (-1)^F e^{-\beta(E-\frac{R_1+R_2}{2})-\beta(a_1h_1+a_2h_2+a_3h_3)-\beta \mu \frac{R_2-R_1}{2}} \right] \ ,
\ee
where $H_{Q}$ is the subspace of the Hilbert space in radial quantization that is annihilated by $Q$ and its conjugate $Q^\dagger$.
The chemical potentials $\beta,a_1,a_2,a_3,\mu$ (together with the constraint $a_1+a_2+a_3=0$) are introduced for the combinations of Cartan generators that commute with $Q$.
$F$ is the fermion number operator and we can take $F=2h_1$ by the spin statistics theorem.
This index at generic chemical potentials respects only a $\mathfrak{su}(1|1)$ subalgebra generated by $Q$ and $Q^\dagger$.

It is often convenient to rephrase the superconformal index as
\be
\cI = \tr_{H_Q} (-1)^F \prod_{j=1}^3 q_j^{h_j+\frac{R_1+R_2}{2}} p^{R_2-R_1} \ ,
\ee
where we have defined
\be
q_j =  e^{-\beta\omega_j}  \qquad p =  e^{-\beta \mu} \ ,
\ee
together with $\omega_j =  1+a_j$. In moving between the two expressions, we have used that states in $H_Q$ saturate the bound~\eqref{bound}. For convergence, we will assume that $|p|<1$, $|q_j|<1$.

In this paper, we will consider an unrefined limit of the superconformal index obtained by specializing the parameters as follows 
\be
\mu \to \frac{1}{2}(\omega_1+\omega_2-\omega_3)
\ee
or equivalently $p \to ( q_1q_2 / q_3 )^{1/2}$. This limit was first considered in \cite{Kim:2013nva} (see also \cite{Beem:2014kka}).
In this limit the index has an enhanced supersymmetry by a second supercharge that we denote $Q' \equiv Q_{++-}^{+-}$ and its conjugate.
The remaining combinations of Cartan generators appearing in the superconformal index commute with the extra supercharges. This leads to dramatic simplifications due to additional cancellations between bosons and fermions. It is straightforward to see that the index simplifies to
\be
\cI = \tr_{\cH_{Q,Q'}} (-1)^F q^{E-R_1} s^{h_1+R_2}  \ ,
\ee
where we have defined $q \equiv q_3$ and $s \equiv q_1 / q_2$. The trace is now over the Hilbert space $H_{Q,Q'}$ of states annihilated by the supercharges $Q$ and $Q'$ and their conjugates in radial quantization. It is clear that the plane rotated by $h_3$ now plays a distinguished role. Indeed, as shown in reference~\cite{Beem:2014kka} the superconformal index in this limit can be interpreted as a vacuum character of a chiral algebra on this plane. For this reason, we will refer to it as the `chiral algebra' limit.

\begin{table}[h]
\begin{center}
  \begin{tabular}{ | c | c | c | c | c | c | c |}
    \hline
    $X$ & $h_1$ & $h_2$ & $h_3$ & $R_1$ & $R_2$ & \\ \hline
    $\phi$ & 0 & 0 & 0 & 1 & 0 & $p^{-1} \sqrt{q_1 q_2 q_3}$ \\ \hline
    $\phi$ & 0 & 0 & 0 & 0 & 1 & $p \sqrt{q_1 q_2 q_3}$ \\ \hline
    $\psi_{++-}^{++}$ & $\frac{1}{2}$ & $\frac{1}{2}$ & -$\frac{1}{2}$ & $\frac{1}{2}$ & $\frac{1}{2}$ & $-q_1q_2$ \\ \hline
    $\psi_{+-+}^{++}$ & $\frac{1}{2}$ & -$\frac{1}{2}$ & $\frac{1}{2}$ & $\frac{1}{2}$ & $\frac{1}{2}$ & $-q_1 q_3$ \\ \hline
    $\psi_{-++}^{++}$ & -$\frac{1}{2}$ & $\frac{1}{2}$ & $\frac{1}{2}$ & $\frac{1}{2}$ & $\frac{1}{2}$ & $-q_2 q_3$ \\ \hline
    $\del \psi $ & $\frac{1}{2}$ & $\frac{1}{2}$ & $\frac{1}{2}$ & $\frac{1}{2}$ & $\frac{1}{2}$ & $q_1q_2q_3$ \\ \hline
 \end{tabular}
\end{center}
\caption{The abelian tensormultiplet has a scalar $\phi$ in fundamental representation of $\mathfrak{so}(5)_R$, 16 fermions $\psi^{R_1R_2}_{h_1,h_2,h_3}$ with $h_1h_2h_2<0$ and a self-dual three-form flux $H$. Recalling that $E(\phi) = 2$, $E(\psi) = 5 / 2$ and $E(H)=3$, the fields commuting with the supercharges $Q$ and $Q^\dagger$ and their contributions to the index are shown above. There is also a contribution from a fermionic equation of motion.}
\label{free}
\end{table}

Let us consider a simple example: the free tensormultiplet in six dimensions. The superconformal index in this case can be evaluated by first enumerating the single letter contributions, which are summarized in table~\ref{free}. The superconformal index is then given by
\bea
\cI & = \PE{ \frac{\left(p+p^{-1}\right) \sqrt{q_1 q_2 q_3}+q_1 q_2 q_3-\left(q_1 q_2+q_2 q_3+q_1 q_3\right)}{\left(1-q_1\right) \left(1-q_2\right) \left(1-q_3\right)} } \ , \\
\eea
where we use the standard definition of the Plethystic exponential. The denominator factors inside the Plethystic exponential come from summing the action of holomorphic derivatives on the single letter contributions. In the chiral algebra limit, the result simplifies to
\be
\cI = \PE{\frac{q}{1-q}} = \prod_{j=1}^\infty \frac{1}{1-q^j}\, .
\ee
In this limit, the index receives contributions only from the scalar $\phi$ corresponding to highest weight in the fundamental of $\mathfrak{so}(5)_R$ i.e. $(R_1,R_2)=(1,0)$ and its holomorphic derivatives in the plane rotated by $h_3$. In particular, the index is independent of $s$. The index is proportional to the vacuum character of the of a free boson in two dimensions, or the vacuum character of $\widehat{\mathfrak{u}}(1)$.


\subsection{$S^5$ Partition Function}\label{subsec:spherepf}

For the interacting $(2,0)$ superconformal theories we do not have a free quantum field theory description and another method must be found to compute the superconformal index. In this subsection, we will summarize the conjecture that the superconformal index of the $(2,0)$ theory of type $A_{N-1}$ is captured exactly by a path integral of the maximal supersymmetric Yang-Mills theory on $S^5$.

The first claim is that the superconformal index in six-dimensions can be expressed as a path integral on $S^1 \times S^5$ with periodic boundary conditions. The chemical potentials $\{\omega_j,\mu\}$ of the superconformal index are now reinterpreted as squashing parameters for the geometry of $S^5$ and expectation values of background R-symmetry gauge fields on $S^5$. The radius of $S^1$ is simply $\beta$. 

The second claim is that this supersymmetric partition function on $S^1 \times S^5$ is captured exactly by the partition function of $\cN=2$ SYM on $S^5$ with gauge group $SU(N)$ and gauge coupling $g^2 = 2 \pi \beta$. This theory is non-renormalizable in five dimensions but is expected to have a UV completion by the $(2,0)$ theory of type $A_{N-1}$ on a circle of radius $\beta$. Although the UV completion may involve new and unknown degrees of freedom, the claim is that the perturbative and non-perturbative spectrum of $\cN=2$ SYM theory on $S^5$ is sufficient to capture all of the protected states contributing to the superconformal index. 

Let us explain in more detail how the chemical potentials $\{ \beta , \omega_j, \mu\}$ of the superconformal index are identified with parameters of the $S^5$ partition function. 
\begin{itemize}
\item The 5d gauge coupling is $g^2 = 2 \pi \beta$. 
\item $\omega_j$ are squashing parameters for the $S^5$ metric - see the final equation of appendix (\ref{eq-squashed-metric}). 
\item $\mu$ is a real mass parameter for the adjoint hypermultiplet inside the $\cN=2$ vectormultiplet.
\end{itemize}
For generic values of the parameters, the $S^5$ partition function preserves two supercharges, $Q$ and $Q^\dagger$, which are identified with those used in the 6d superconformal index.

Finally, we note that the transformation between the superconformal index and the partition function on $S^1\times S^5$ will likely involve an anomalous background coordinate and $R$-symmetry gauge transformation. Thus we can expect them to agree up to a multiplicative factor determined by global anomalies. More precisely, we will find that
\be
Z_{S^5} = e^{-\cF} \, \mathcal{I}
\ee
where $\cF$ is a finite Laurent polynomial in the parameters $\beta \omega_1,\ldots,\beta \omega_3$ and $\beta m$ that can be determined from the anomaly polynomial of the six-dimensional theory~\footnote{It will be demonstrated in forthcoming work that the finite Laurent polynomial $\cF$ is an equivariant integral of the anomaly polynomial~\cite{Bobev:2015kza}.}. We will confirm this structure in examples.


\subsection{Computation of $S^5$ Partition Function}\label{subsec:comp}

The partition function $Z_{S^5}$ can be evaluated exactly using the method of supersymmetric localization or alternatively the refined topological string partition function \cite{Hosomichi:2012ek,Kallen:2012cs,Kallen:2012va,Kim:2012ava,Imamura:2012xg,Lockhart:2012vp,Imamura:2012bm,Kim:2012qf}. Here we focus on the former approach. A short review of the localization computation is given in Appendix \ref{sec-S5-partitionftn}.

The path integral localizes to a matrix integral over a set of saddle points. The saddle points are classified as follows. Firstly, one of the 5 adjoint scalars has a constant non-zero vacuum expectation value $\la \phi \ra = ia$. This is the real scalar in the $\cN=1$ vectormultiplet. In addition, there are non-perturbative instanton saddle points on top of this background. The instanton saddle points are the self-dual Yang-Mills instantons on $\mathbb{CP}^2$ base of the Hopf fibration $S^5 \to \mathbb{CP}^2$. The 1-loop determinant of the fluctuations around these saddle points gives rise to perturbative and non-perturbative contributions to the measure of the matrix integral.

Our crucial assumption is that the one-loop and non-perturbative contributions factorize into contributions from 3 fixed circles of the $U(1)^3$ isometry group of the squashed $S^5$. We will denote these fixed circles by $S^1_{(i)}$ with $i=1,2,3$. This factorization can be verified explicitly for the perturbative contributions but has not been demonstrated conclusively for the non-perturbative contributions. Under this assumption, the partition function has the form \cite{Lockhart:2012vp,Kim:2012qf}:
\be
    Z_{S^5}(m,\vec \omega,\tau) = \int [da] \ e^{\frac{2\pi^2}{\beta \omega_1\omega_2\omega_3}(a,a)} \prod_{i=1}^3 Z^{(i)} \,,
\ee
where as above $\la \phi \ra = ia$ is the $\cN=1$ vectormultiplet scalar expectation value and $\beta = g^2 / 2\pi$.
The measure of integration is
\be
	[da] = \frac{i^{N-1}}{N!} d^{N-1}a \ ,
\ee
and the integral domain is over $i\mathbb{R}^{N-1}$. 
The contributions $Z^{(i)}$ are copies of the 5d Nekrasov partition function $Z_{\rm Nek}(a,m,\epsilon_1,\epsilon_2)$ on $S^1 \times \mathbb{R}^4_{\ep_1,\ep_2}$ where the circle $S^1$ has radius $r$. The equivariant parameters $(r,m,\epsilon_1,\epsilon_2)$ at each fixed circle are replaced as shown in table~\ref{sphere}. 

\begin{table}[h]\label{tb:parameters}
\begin{center}
\begin{tabular}{ | l | c | c | c | c | }
  \hline                       
  \, & $r$ & $\epsilon_1$ & $\epsilon_2$ & $m$ \\ \hline
  $S^1_{(1)}$ & $2\pi / \omega_1$ & $\omega_2$ & $\omega_3$ & $\mu + \frac{3}{2}\omega_1$ \\
   $S^1_{(2)}$ & $2\pi/\omega_2$ & $\omega_3  $ & $\omega_1 $ & $\mu + \frac{3}{2}\omega_2$ \\
   $S^1_{(3)}$ & $2\pi/\omega_3$ & $\omega_1  $ & $\omega_2  $ & $\mu + \frac{3}{2}\omega_3$ \\
  \hline  
\end{tabular}
\end{center}
\caption{\emph{The $S^5$ partition function is constructed from three copies of the 5d Nekrasov partition function with the parameters identified as above. The arguments in the final three columns may be taken modulo $\omega_j$ in the row corresponding to $S^1_{(j)}$}}
\label{sphere}
\end{table}

The Nekrasov partition function $Z_{\rm Nek}(a,m,\epsilon_1,\epsilon_2)$ on $S^1 \times \mathbb{R}^4_{\epsilon_1,\epsilon_2}$ can be expressed as a supersymmetric index with both perturbative contributions from fundamental BPS particles and non-perturbative contributions from BPS instanton particles \cite{Nekrasov:2002qd,Nekrasov:2003rj}. It is constructed from moduli space of $k$ $U(N)$ instantons $\cM_{k,N}$ which carries an action of $U(1)^2 \times U(N)$ corresponding to rotations in two orthogonal planes of $\mathbb{R}^4$ and gauge transformations. The parameters $\epsilon_1, \epsilon_2$ and $\vec a = (a_1,\ldots,a_N)$ are equivariant parameters for these symmetries. It is well-known that the fixed points with respect to this action are labelled by an $N$-tuple of Young tableaux $\vec Y = (Y_1,\ldots,Y_N)$ such that the total number of boxes is given by $| \vec Y | = k$.

The starting point for computing the partition function is the equivariant Chern character of the universal bundle $\mathcal{E}$ evaluated at the fixed point corresponding to $\vec{Y}$ \cite{Losev:2003py,Shadchin:2004yx}:
\bea
    {\rm Ch}_{\vec{Y}}(\mathcal{E}) &= \mathcal{W} - (1-e^{-\epsilon_1})(1-e^{-\epsilon_2})e^{\epsilon_+} \mathcal{V} \,,
\eea
where we define $\epsilon_+ \equiv ( \epsilon_1+\epsilon_2 ) / 2$ and 
\be
    \mathcal{V} = \sum_{I=1}^N \sum_{(i,j)\in Y_I} e^{a_I -(i-\frac{1}{2})\epsilon_1 - (j-\frac{1}{2})\epsilon_2} \,, \quad \mathcal{W} = \sum_{I=1}^N e^{a_I} \,.
\ee
Then the equivariant index of the tangent bundle $\mathcal{TM}_{k,N}$ at the critical point labelled by $\vec{Y}$ is given by \cite{Losev:2003py,Shadchin:2004yx}
\bea\label{eq-eqidx-tangent}
    {\rm Ind}_{\mathcal{TM}_{\vec{Y}}} &= - \frac{ {\rm Ch}_{\vec{Y}}(\mathcal{E}){\rm Ch}_{\vec{Y}}(\mathcal{E}^*) }{(1-e^{-\epsilon_1})(1-e^{-\epsilon_2})} \\
    &= - \frac{ \mathcal{W}\,\mathcal{W}^* }{(1-e^{-\epsilon_1})(1-e^{-\epsilon_2})} + e^{\epsilon_+} (\mathcal{V}\,\mathcal{W}^* + \mathcal{V}^*\,\mathcal{W}) - (1-e^{\epsilon_1})(1-e^{\epsilon_2}) \mathcal{V}^* \, \mathcal{V} \,,
\eea
where the conjugate $*$ flips the signs in the exponents. This amounts to the equivariant index for the $U(N)$ vector multiplet. The first term in the second line is independent of the instanton number (independent of $\mathcal{V}$) and thus is regarded as the perturbative contribution of the vector multiplet. The other terms are the instanton contributions coming from the zero modes in the instanton background. 

For the maximal SYM theory, there are also contributions from the hypermultiplet in the adjoint representation of the gauge group. The hypermultiplet in the instanton background has the equivariant index of the form
\be
    {\rm Ind}_{\mathcal{V}_{\vec{Y}}^H} = e^{ m -\epsilon_+ } \frac{ {\rm Ch}_{\vec{Y}}(\mathcal{E})\,{\rm Ch}_{\vec{Y}}(\mathcal{E^*}) }{(1-e^{-\epsilon_1})(1-e^{-\epsilon_2})} \,,
\ee
where $m$ is the equivariant parameter for the flavor symmetry of the hypermultiplet. As for the vector multiplet, this index can also be divided into the perturbative contribution, the term without $\mathcal{V}$, and the instanton contribution, the other terms.

We use the conversion rule from the equivariant index to the partition function at the fixed point
\be
    {\rm Ind} = \sum_i n_i e^{\omega_i} \quad \rightarrow \quad \prod_i \omega_i^{-n_i} \,,
\ee
where the integer $n_i$ denotes the degeneracy for given weight $\omega_i$. Also, we should keep in mind the momentum factor $\sum_{t\in \mathbb{Z}} e^{\frac{2\pi}{r} t}$ along the temporal circle in five-dimensions. It should be put on each equivariant index. The instanton partition function then becomes
\be
    Z_{\rm inst}^{U(N)} = \sum_{\vec{Y}} \mathfrak{q}^{|\vec{Y}|} \prod_{i,j=1}^N \prod_{s\in Y_i}\frac{ \sin\frac{r(E_{ij}(s)+m-\epsilon_+) }{2}\sin\frac{r(E_{ij}(s)-m-\epsilon_+ )}{2} }{ \sin\frac{r E_{ij}(s)}{2}\sin\frac{r (E_{ij}(s) -2\epsilon_+) }{2} } \,,
\ee
where $s$ denotes a position of a box in the Young tableau $Y_i$ and
\be
    E_{ij}(s) = a_i - a_j - \epsilon_2 h_i(s) + \epsilon_1( v_j(s) + 1 ) \,.
\ee
 $h_i(s)$ and $v_j(s)$ are the distance from $s$ to the right and bottom end of $i$-th and $j$-th Young tableaux, respectively. $\mathfrak{q}=e^{-\frac{4\pi^2r}{g^2}}$ is the instanton fugacity.

The perturbative contribution can be computed in the similar manner. 
However, there is an ambiguity in computation of the perturbative partition function 
associated to boundary conditions of $\mathbb{R}^4$ at infinity.
Studying the correct boundary conditions goes beyond the scope of this paper. Here we simply take the average of the equivariant index with its charge conjugation and compute the perturbative contribution, which makes the equivariant index invariant under the charge conjugation. This choice is rather convenient for seeing the simplifications that occurs in the chiral algebra limit at the level of the Nekrasov partition function.
We also need to regularize the infinite products involved in the perturbative contribution. To regularize them, we shall use Barnes' multiple gamma functions defined in~\cite{2003math......6164N} as follows:
\begin{equation}
	\Gamma_N(z|w_1,\cdots,w_N) = \mathrm{exp}\left( \partial_s\zeta_N(s,z;w_1,\cdots,w_N)|_{s=0} \right) \sim \prod_{n_1,\cdots,n_N\ge0} (z+n\cdot w)^{-1} \,,
\end{equation}
and Barnes' zeta functions defined by the series
\begin{equation}
	\zeta_N(s,z|w_1,\cdots,w_N) = \sum_{n_1,\cdots,n_M \ge 0}(z+n\cdot w)^{-s} \,.
\end{equation}
Therefore the regularized perturbative partition function is given by
\bea\label{eq:pert-part}
    Z_{\rm pert} &= \prod_{e\in\Delta} \left[\frac{ \tilde\Gamma_3\!\left((e,a)\pm m+\epsilon_+\right) }
     { \tilde\Gamma_3'\!\left((e,a)\right)\tilde\Gamma_3\!\left((e,a)+\frac{2\pi}{r}+2\epsilon_+\right) } \right]^{1/2} \,,
\eea
We have defined $\tilde\Gamma_3(z)\equiv \Gamma_3(z;\frac{2\pi}{r},\epsilon_1,\epsilon_2)\Gamma_3(1-z;\frac{2\pi}{r},-\epsilon_1,-\epsilon_2)$ and put the prime on it to deal with the zero modes at $z=0$ such that $\tilde\Gamma'(0) =\lim_{z\rightarrow 0} z\tilde\Gamma(z)$.

It is often convenient to rewrite the perturbative contribution in the more concise expression
\be
    Z_{\rm pert} \sim \prod_{e \in \Delta } \left[\frac{ (e^{r(e,a)};p,q)'_\infty(e^{r(e,a)}pq;p,q)_\infty }{ (e^{r(e,a)\pm m}\sqrt{pq};p,q)_\infty  }\right]^{1/2} \,,
\ee
up to regularization factors.
Here, $(x;p,q)_\infty\equiv\prod_{n,m=0}^\infty(1-xp^nq^m)$ is the multiple $q$-Pochhammer symbol and $p\equiv e^{-r\epsilon_1}, q\equiv e^{-r\epsilon_2}$ and the prime denotes the zero modes at $x=0$ are absent. 

As explained above, the chiral algebra limit of the 6d superconformal index is approached by tuning the real mass parameter $\mu \to \frac{1}{2}(\omega_1+\omega_2-\omega_3)$ \cite{Kim:2013nva,Beem:2014kka}. Using that the contributions from the circle $S^1_{(j)}$ are periodic in $\omega_j$ we can deduce that from the point of view of each fixed point this is equivalent to the limits
\be
(1) \quad m = \pm \epsilon_-  \,,\qquad (2) \quad m = \pm \epsilon_+ \ ,
\ee
where $\epsilon_\pm = (\pm\epsilon_1+\epsilon_2)/2$. These points correspond to an enhancement of supersymmetry and lead to additional cancellations. 

First consider the limit $m=\pm \epsilon_-$. We note that there is a universal center of mass factor multiplied to each non-zero instanton number contribution taking the form of
\be
\frac{\sin\frac{r(\epsilon_-+m)}{2}\sin\frac{r(\epsilon_--m)}{2}}{\sin\frac{r\epsilon_1}{2}\sin\frac{r\epsilon_2}{2}} \ ,
\ee
and it vanishes in both cases $m=\pm \epsilon_-$. Therefore the instanton contribution is unity in this limit. Furthermore, the perturbative contributions simplifies to a product of sine functions, so that
\be
Z(a,m=\pm\epsilon_-,\epsilon_1,\epsilon_2) = \left(\frac{r}{2\pi}\right)^{N/2}\prod_{e>0} 2\sin(\frac{r}{2}(e,a))\ ,
\ee
which we get from (\ref{eq:pert-part}) using the identities of $\Gamma_3$.
This formula is derived for $U(N)$. For $U(1)$ the perturbative and instanton contributions are trivial. Thus we expect that the answer for $SU(N)$ is simply obtained by imposing the condition $a_1+\cdots + a_N=0$ on the above formula.

Second, consider the limit $m = \pm\epsilon_+$. In this case, the contributions to the index from the vectormultiplet and adjoint hypermultiplet cancel out exactly. Thus the perturbative contribution is unity and the instanton contribution simply counts the number of fixed points. Therefore we have
\be
Z^{U(N)}_{\mathrm{inst}}(a,m = \epsilon_+,\epsilon_1,\epsilon_2) = \sum_{\vec{Y}} \mathfrak{q}^{|\vec{Y}|} = \prod\limits_{n=1}^{\infty} \frac{1}{(1-\mathfrak{q}^n)^N } \, .
\ee
In this simple limit, we expect that the $SU(N)$ answer is obtained by dividing by the partition function for $U(1)$ so that 
\be
Z^{SU(N)}_{\mathrm{inst}}(a,m = \epsilon_+,\epsilon_1,\epsilon_2) = \prod\limits_{n=1}^{\infty} \frac{1}{(1-\mathfrak{q}^n)^{N-1} } \ .
\ee


\subsection{Codimension 4 Defects}\label{subsec:codim4}

It is expected that the six-dimensional $(2,0)$ theories have codimension four surface defects labelled by irreducible representations of $\mathfrak{g}=A_{N-1}$ with highest weight $\lambda$. Upon dimensional reduction on $S^1$ they should become supersymmetric Wilson loops in $\cN=2$ SYM with gauge group $SU(N)$ in the same irreducible representations.
Supersymmetric Wilson loops in five dimensions have been studied in \cite{Young:2011aa,Kim:2012qf,Assel:2012nf}.

Let us begin with an abelian tensormultiplet in six-dimensions where there is an explicit description of the codimension four defects as Wilson surfaces. In order to preserve $Q = Q^{++}_{---}$ we wrap the Wilson surface around the M-theory circle $S^1$ and one of the three closed circles $S^1_{(j)} \subset S^5$. Then we construct the Wilson surface in the euclidean spacetime
\be
    \exp\bigg( n \int_{S^1 \times S^1_{(j)}} (iB + \Phi \, d\tau \!\wedge\! ds_j) \bigg) \,,
\ee
where $n \in \mathbb{Z}$ is the abelian charge and $d\tau$ and $ds_j$ are line elements along $S^1$ and $S^1_{(j)}$ respectively. $B$ denotes the two-form gauge field with self-dual curvature $H = dB$ and $\Phi$ is one of the five scalars in the abelian tensormultiplet. The scalar $\Phi$ is characterized as a singlet under the generators $R_1$ and $R_2$. Note that the Dirac quantization in 6d CFT forces the charge $n$ to be an integer.

Turning off all of the chemical potentials in the superconformal index, it is straightforward to check that the supersymmetric Wilson surface supported at the fixed point $S^1_{(j)}$ always preserves the two supercharges
\be
Q_{---}^{++} \,, \quad Q_{---}^{--} \ ,
\ee  
together with six more depending on the fixed point inserted and all of their conjugates. For instance, in the case of $S^1_{(1)}$ the six additional supercharges preserved are 
\be
Q_{-++}^{++}\,, \quad Q_{-++}^{--}\,, \quad Q_{++-}^{+-}\,, \quad Q_{++-}^{-+}\,, \quad Q_{+-+}^{+-}\,, \quad Q_{+-+}^{-+} \ .
\ee 
Those preserved at the remaining fixed points are obtained by symmetric permutations of the lower indices.

Turning on the general chemical potentials in the superconformal index, the Wilson surface respects only $Q = Q_{---}^{++}$ and its conjugate. This means we can still define and compute the superconformal index with generic chemical potentials in the presence a Wilson surface on any $S^1 \times S^1_{(j)}$. However, the Wilson surface commutes with the second supercharge $Q' \equiv Q_{++-}^{+-}$ preserved in the chiral algebra limit only if it wraps $S^1_{(1)}$ or $S^1_{(2)}$. From the six-dimensional perspective, these cases correspond to codimension four defects transverse to the chiral algebra plane. We will concentrate only on these cases in the present work.

The dimensional reduction of the Wilson surface operators along the $S^1$ gives rise to the supersymmetric Wilson loops in the five-dimensional $U(1)$ gauge theory on $S^5$. In particular, the two-form $B$ provides the five-dimensional gauge field $A_\mu \equiv \beta B_{\tau\mu}$ where $ \mu=1,\ldots,5$ and similarly we define the five-dimensional scalar $\phi = \beta \Phi$. Taking these fields independent of $\tau$ and integrating over the M-theory circle $S^1$, we find a supersymmetric Wilson loop in five dimensions
\be
    \exp\bigg( n \oint_{S^1_{(j)}} (iA + \phi \, ds_j) \bigg) \ .
\ee
As in 6d abelian case, the charge $n$ is quantized, which naively seems not to be the case since the abelian theory has no charged object perturbatively. However, non-perturbative objects, for example the singular instantons which we assume to be involved in this paper, can carry nontrivial charge and thus the quantization condition is required.

In the case of the six-dimensional $(2,0)$ theory of type $\mathfrak{g} = A_{N-1}$ we cannot formulate the codimension four defect directly on $S^1 \times S^1_{(j)}$. Nevertheless, taking inspiration from the abelian tensormultiplet, we can conjecture that it is computed by inserting a supersymmetric Wilson loop on $S^1_{(j)}$,
\be
 \tr_{\lb} \, \mathrm{P} \, \exp\bigg( \oint_{S^1_{(j)}} (iA + \phi \, ds_j) \bigg) \ ,
\ee
where the trace is taken in the representation of $SU(N)$ of highest weight $\lb$.

On the saddle points we have $A=0$ and $\phi = i a$ and hence integrating over $S^1_{(j)}$ of length $2\pi / \omega_j$ the supersymmetric Wilson loop will make a classical contribution $\tr_{\lb} (e^{2\pi i a /\omega_j})$ to the integrand of the partition function. The presence of the supersymmetric Wilson loop does not affect the 1-loop perturbative computation. However, the Wilson loop in general receives instanton corrections. 

The Wilson loop partition function on $S^1\times \mathbb{R}^4$ is obtained by inserting the equivariant Chern character $Ch_{\vec{Y}}(\mathcal{E})$ of the universal bundle to the instanton partition function.
For example, in the case of a fundamental Wilson loop, we have the expectation value of the form
\be
W_{\rm fund}^{U(N)} = \frac{1}{Z_{\rm inst}^{U(N)}} \sum_{\vec{Y}} \mathfrak{q}^{|\vec{Y}|} Ch_{\vec{Y}}(\mathcal{E})\prod_{i,j=1}^N \prod_{s\in Y_i}\frac{ \sin\frac{r(E_{ij}(s)+m-\epsilon_+) }{2}\sin\frac{r(E_{ij}(s)-m-\epsilon_+ )}{2} }{ \sin\frac{r E_{ij}(s)}{2}\sin\frac{r (E_{ij}(s) -2\epsilon_+) }{2} } \ ,
\ee
normalized by the bare partition function.
To insert a Wilson loop on the circle $S^1_{(j)}$, we insert this factor with  parameters at $S^1_{(j)}$ to the partition function.

Let us again consider the special limits $m = \epsilon_+$ and $m = \epsilon_-$. Firstly the Wilson loop in the limit $m=\epsilon_+$ receives a rather simple instanton correction. After the huge cancellations, we obtain
\be\label{eq-Wilson-U(N)}
    W_{\rm fund}^{U(N)}(a,m=\epsilon_+,\epsilon_1,\epsilon_2) = \sum_{i=j}^N e^{ir a_j} \ {\rm PE}\left[-(1-e^{-ir\epsilon_1})(1-e^{-ir\epsilon_2})\frac{\mathfrak{q}}{1-\mathfrak{q}}\right] \ .
\ee
However, we do not discuss Wilson loops in this limit further in this work since it corresponds to inserting the Wilson loop in the chiral algebra plane.

On the other hand when $m=\epsilon_-$, since the center of mass factor multiplied to each instanton sector vanishes, the Wilson loop reduces to its classical value
\be
     W_{\rm fund}^{U(N)}(a,m=\epsilon_-,\epsilon_1,\epsilon_2)= \sum_{i=j}^N e^{ir a_j} \ .
\ee
We conjecture that the $U(N)$ fundamental Wilson loop involves the $U(1)$ Wilson loop factor and thus the $SU(N)$ Wilson loop expectation value is given by
\be
    W^{SU(N)}_{\rm fund} = W_{\rm fund}^{U(N)} / \, W^{U(1)} \ ,
\ee
where $W^{U(1)}$ is the abelian Wilson loop of unit charge under the overall $U(1)$ gauge group. One can also identify this abelian Wilson loop expectation value from the $U(1)$ gauge theory computation. The answer is
\be\label{eq-Wilson-U(1)}
    W^{U(1)}=  e^{ir a}\ {\rm PE}\left[(1-e^{-ir\epsilon_1})(1-e^{-ir\epsilon_2}) \  z_{U(1)} \right] =  e^{ir a}\ {\rm PE}\left[e^{-ir\epsilon_1}(1-e^{ir(\epsilon_-\pm m)})\frac{\mathfrak{q}}{1-\mathfrak{q}} \right] \ ,
\ee
where $a$ is the equivariant parameter of the $U(1)$ gauge symmetry and $z_{U(1)}$ is the letter index of the $U(1)$ instanton partition function given by
\be
    z_{U(1)} = \frac{\sin\frac{r(\epsilon_-+m)}{2}\sin\frac{r(\epsilon_--m)}{2}}{\sin\frac{r\epsilon_1}{2}\sin\frac{r\epsilon_2}{2}} \frac{\mathfrak{q}}{1-\mathfrak{q}} \ .
\ee
This abelian Wilson loop shows the expected behavior in the special limits. In the limit $m=\epsilon_+$, it reproduces the Plethystic exponential term in (\ref{eq-Wilson-U(N)}) and, in the second limit $m=\epsilon_-$, it vanishes as expected.
This conjecture for the $SU(2)$ fundamental Wilson loop can also be checked by comparing against the Wilson loop computation value of the $Sp(1)$ gauge theory, which we have confirmed for the single instanton calculus.

The abelian Wilson loop can be interpreted as a heavy tensor multiplet in the 6d theory. The $U(1)$ instanton partition function is the Plethystic exponential of the letter index $z_{U(1)}$ and agrees with the index of a single tensor multiplet on $T^2 \times \mathbb{R}^4$~\cite{Kim:2011mv}. The index of the heavy tensor fields can be obtained from the index of the single tensor multiplet by removing the factors corresponding to the motion along $\mathbb{R}^4$. We can achieve it by multiplying the factor $(1-e^{-ir\epsilon_1})(1-e^{-ir\epsilon_2})$ to the letter index $z_{U(1)}$. Thus the index of a heavy 6d tensor multiplet is the precisely the index of the abelian Wilson loop in (\ref{eq-Wilson-U(1)}).

The Wilson loop expectation value in the other representations can be computed in the similar manner. We have to insert the corresponding Chern character to the instanton partition function.
One can construct the equivariant Chern character in the general representation of the gauge group using that of the universal bundle $\mathcal{E}$. For instance, the equivariant characters in the symmetric and anti-symmetric representations are given by \cite{Shadchin:2004yx}
\bea
    Ch_{\rm sym} &= \frac{1}{2}\left[ Ch_q(\mathcal{E})^2 + Ch_{q^2}(\mathcal{E}^2) \right] \ ,\\
    Ch_{\rm anti} &= \frac{1}{2}\left[ Ch_q(\mathcal{E})^2 - Ch_{q^2}(\mathcal{E}^2) \right] \ .
\eea
Here $q$ stands for the exponential of equivariant parameters and thus $Ch_{q^2}(\mathcal{E})$ means the equivariant character of $\mathcal{E}$ with doubled equivariant parameters. We can apply the similar construction for other representations. 
The instanton partition functions with the insertion of these characters yield the corresponding Wilson loop partition functions.

The Wilson loop partition functions in general representations have the universal center of mass factor which vanishes in the limit $m=\epsilon_-$. Thus, in this limit, the Wilson loop partition function in the representation labelled by highest weight $\lambda$ takes the form of the classical expectation value
\be
    W^{U(N)}_\lambda (a,m=\epsilon_-,\epsilon_1,\epsilon_2) = {\rm Tr}_\lambda (e^{ira}) \ .
\ee
On the other hand, it exhibits a rather complicated expression when we take the limit $m=\epsilon_+$ of Wilson loop in a general representation, which we do not discuss in this paper.


\subsection{Codimension 2 Defects}\label{subsec:codim2}

The non-abelian $(2,0)$ theories are expected to have codimension 2 surface defects whose study was initiated in~\cite{Gaiotto:2009we,Gaiotto:2009hg} in the context of the construction of four dimensional theories of class $\cS$. For a detailed discussion of the classification and properties of codimension two defects see~\cite{Gukov:2006jk,Gomis:2007fi,Alday:2009fs,Alday:2010vg,Chacaltana:2012zy}. 

For $\mathfrak{g}=A_{N-1}$ the codimension 2 defects are in 1-1 correspondence with homomorphisms $\rho : \mathfrak{su}(2) \to \mathfrak{g}$. They can be labelled by a partition $[n_1,\ldots,n_s]$ with $\sum_{j=1}^s n_j = N$ and by convention we take $n_i \leq n_j$ if $i<j$. This data encodes how the fundamental representation decomposes $N \to n_1 + \cdots n_s$ into representations of the image of $\rho$.  An important property of codimension two defects is that they support a flavor symmetry. Let $\ell_j$ be the number of times that the number $j$ appears in the partition $[n_1,\ldots,n_s]$ so that $\sum_{j} j \ell_j=N$. Then the flavor symmetry supported by the defect is $\mathfrak{s}( \oplus_j \, \mathfrak{u}(\ell_j))$.

Codimension 2 defects can be understood as transverse M5 branes intersecting with the primary stack of $N$ coincident M5 branes. However, there is an alternative description of codimension 2 defects discussed in~\cite{Tachikawa:2011dz,Kanno:2011fw} whose connection to our computations is more transparent. This involves the primary stack of $N$ M5-branes probing a multi-centered Taub-NUT (TN) space with $s$ singularities. The M5-branes wrap the circle fiber and extend along a radial direction of the base $\mathbb{R}^3$ of TN, with a number $n_i$ M5 branes ending on $i$-th singularity. Thus the data classifying such configurations is a partition $[n_1,\cdots,n_s]$ of $N$. Each set of $n_i$ M5-branes is supported on a cigar in the TN geometry. When all $s$ centers coincide, TN develops an $\mathbb{Z}_s$ orbifold singularity. Such configuration generates a codimension 2 defect at the tip of the cigar spanned by the M5 branes. Later in this section, we will compute the partition function in the presence of the codimension 2 defects using the instanton calculus of the 5d SYM on the $\mathbb{Z}_s$ orbifold plane.

The brane descriptions indicate the symmetries that are preserved by a codimension 2 defect. Firstly, the six-dimensional conformal and R-symmetries are broken to
\bea
\mathfrak{so}(2,6) & \to \mathfrak{so}(2,4) \oplus \mathfrak{so}(2)_1 \ , \\
\mathfrak{so}(5) & \to \mathfrak{so}(2)_2 \oplus \mathfrak{so}(3)_R \ .
\eea 
Let us orient the defect such that the unbroken conformal symmetry $\mathfrak{so}(2,4)$ and R-symmetry $\mathfrak{so}(3)_R$ have Cartan generators $(E,h_2,h_3)$ and $R\equiv R_1$ respectively. The remaining $\mathfrak{so}(2)_1$ and $\mathfrak{so}(2)_2$ symmetries are generated by $h_1$ and $R_2$. The combination $r = - h_1 - 2R_2$ generates a superconformal R-symmetry $\mathfrak{u}(1)_r$ while the remaining diagonal combination $d=h_1+R_2$ becomes an additional $\mathfrak{u}(1)_d$ flavor symmetry. The full symmetry preserved by a codimension 2 defect is thus $\mathfrak{su}(2,2|2) \oplus \mathfrak{u}(1)_d$ with the first factor being the $\cN=2$ superconformal algebra of the 4d theory living on the intersection. For the other orientation, similar relations hold with cyclic permutations on $h_1,h_2,h_3$.

Let us consider in detail the case where the codimension 2 defect spans the plane rotated by $(h_2,h_3)$ (we could also consider $(h_1,h_3)$ with similar results) so that in particular it fills the chiral algebra plane rotated by $h_3$. In this case it is illuminating to write the supersymmetry algebra generated by the charges $Q$ and $Q'$ in terms of the $\mathfrak{su}(2,2|2) \oplus \mathfrak{u}(1)$ generators as
\bea
\{ Q , Q^\dagger\} & = E-2R+r-h_2-h_3 \ , \\
\{ Q' ,Q'^\dagger \} & = E - 2R -r +h_2-h_3 \, .
\eea
The chiral algebra limit of the superconformal index can be written
\be
    \cI = \tr_{\cH_{Q,Q'}} (-1)^F q^{E-R} s^{d} \prod_{j}t_j^{f_j} \ ,
\ee
where we have introduced additional fugacities $t_j$ for the Cartan generators $f_j$ of the $\mathfrak{s}( \oplus_j \, \mathfrak{u}(\ell_j))$ flavor symmetry of the defect. Note that $s$ becomes a fugacity for the additional $\mathfrak{u}(1)_d$ flavor symmetry. In what follows, we find that all states contributing to the superconformal index have $d=0$ and therefore the index is independent of $s$. The index then coincides with the Schur limit of the $\cN=2$ superconformal index for the $\mathfrak{su}(2,2|2)$ algebra in four dimensions \cite{Gadde:2011ik,Gadde:2011uv,Gaiotto:2012xa}.  

As before, we cannot compute the superconformal index in the presence of a codimension 2 defect directly in six dimensions. However, it is expected that for a codimension 2 defect wrapping $S^1$, there is an equivalent description as a monodromy defect in 5d $\cN=2$ SYM. Such monodromy defects were introduced in the physics literature in reference~\cite{Gukov:2006jk} in the context of 4d $\cN=4$ SYM. Extrapolating the arguments there to five dimensions, we expect that a monodromy defect labelled by the partition $\rho = [n_1,\ldots,n_s]$ has an alternative description by coupling 5d $\cN=2$ SYM to a 3d $\cN=4$ $\sigma$-model whose target space is $T^*(G / \mathbb{L})$ where $\mathbb{L} = S(U(n_1) \times \cdots U(n_s)))$ is the associated Levi subgroup. 

\begin{figure}[h]
\begin{center}
\includegraphics[width=80mm]{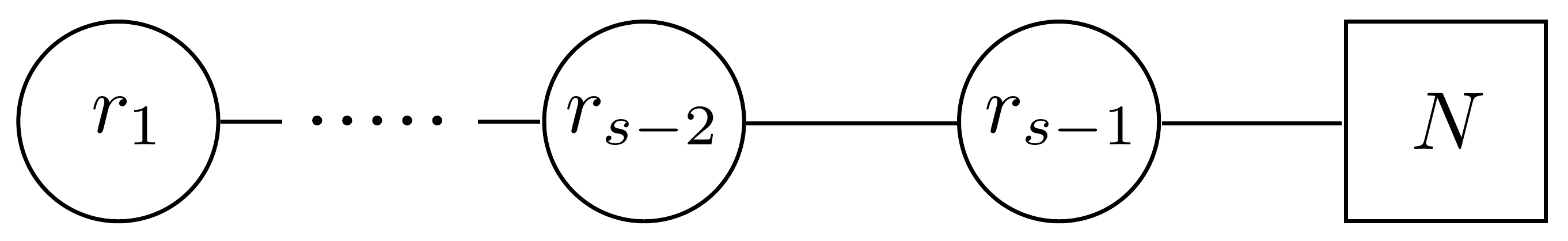}
\end{center}
\vspace{-20pt}
\caption{Linear quiver}
\label{fig:linear}
\end{figure}

This $\sigma$-model has a UV description as the gauge theory $T_{\rho}(\mathfrak{g})$ introduced reference in~\cite{Gaiotto:2008ak}, which is the linear quiver shown in figure \ref{fig:linear}. In particular, there is a sequence gauge groups $U(r_i)$ where $r_i = n_1+\cdots + n_i$ for $i=1,\ldots,s-1$ (so that $r_s=N$). There is an $\mathfrak{su}(N)$ symmetry acting on the $N$ hypermultiplets at the final node, and on $S^3$ we can turn on corresponding real mass parameters in the Cartan subalgebra. Here we work instead with imaginary mass parameters $a = (a_1,\ldots,a_N)$ with $\sum_j a_j=0$. They are identified with the expectation values of the scalar fields in the vectormultiplet of the bulk theory. There is a also topological $\mathfrak{u}(1)^{s-1}$ manifest in the linear quiver description, which is enhanced to $\mathfrak{s}( \oplus_j \, \mathfrak{u}(\ell_j))$ in the infrared. Let us introduce corresponding FI parameters by $m_{j}$ where $\sum_{j} m_j = 0$ dual to the Cartan generators of $\mathfrak{s}( \oplus_j \, \mathfrak{u}(\ell_j))$. They are identified with the flavor fugacities of the 6d superconformal index by $t_{j} = e^{-\beta m_{j}}$.
\\

At this point, the most rigorous way to proceed would be to attempt an exact localization computation for $\cN=2$ SYM on $S^5$ coupled to $T_\rho(\mathfrak{g})$ supported on $S^3 \subset S^5$. This computation is beyond the scope of the present paper. Instead, we will employ an effective description of these surface defects as monodromy defects of Levi type $\mathfrak{l} = \mathfrak{s}(\mathfrak{u}(n_1) \oplus \cdots \oplus \mathfrak{u}(n_s))$ and factorize the computation of the partition function into contributions from three fixed circles of $S^5$. The validity of our procedure will be tested a posteriori by reproducing the $S^3$ partition function of $T_\rho(\mathfrak{g})$ by sending the bulk coupling $g^2 \to 0$.

Let us now describe the computational scheme. Our first assumption is that in the presence of a surface defect of type $\rho$, the partition function can once again be expressed as a matrix integral, whose integrand is factorized into contributions localized at the three fixed circles $S^1_{(a)}$, $a=1,2,3$. For definiteness, suppose that the surface defect is supported on the three-sphere $S^3_{(3)} \subset S^5$ containing $S^1_{(1)}$ and $S^1_{(2)}$ as Hopf linked circles but supported away from $S^1_{(3)}$. Thus contributions localized at $S^1_{(3)}$ should not be changed by the presence of the defect. On the other hand, from the perspective of $S^1_{(1)}$ and $S^1_{(2)}$ the defect is supported on subspaces $S^1 \times \mathbb{C}_{\epsilon_1}$ and $S^1 \times \mathbb{C}_{\ep_2}$ of $S^1 \times \mathbb{C}^2_{\ep_1,\ep_2}$ with the equivariant parameters identified as in Table~\ref{tb:parameters}.

Thus it is reasonable that the partition function of the combined system can be expressed as
\be
 \int [da] \sum_{\sigma=1}^{N_\rho}  e^{\frac{2\pi^2}{\beta \omega_1 \omega_2 \omega_3}(a,a)  -\frac{4\pi^2}{\beta\omega_1\omega_2}(\sigma(\vec{m}),a) } Z^{(1)}_{\rho,\sigma} Z^{(2)}_{\rho,\sigma} Z^{(3)} \, .
\ee
where the contribution from the third fixed point $Z^{(3)}$ is the same as in the absence of the defect. The measure now becomes
\be
	[da] = \frac{i^{N-1}}{n_1!n_2!\cdots n_s!} d^{N-1}a
\ee
as the gauge group is broken to the subgroup $\mathbb{L}$.
The summation $\sigma = 1,\ldots, N_\rho$ runs over the supersymmetric vacua of the three-dimensional theory $T_\rho(\mathfrak{g})$ on $S^1 \times \mathbb{C}$. The number of these vacua is in general $N_\rho = N! / ( n_1!  n_2 ! \ldots n_s!) $.
Note that the classical action has an additional contribution with the monodromy parameter $\vec{m}$ whose derivation on round $S^5$ is given in the appendix~\ref{sec-S5-partitionftn}.

Our second assumption is that the contributions $Z^{\rho,\sigma}_{(j)}$ at fixed points $S^1_{(1)}$ and $S^1_{(2)}$ are given by the 5d Nekrasov partition function in the presence of a monodromy defect. The partition $\rho = [n_1,n_2,\ldots,n_s]$ labelling the surface defect determines the Levi subgroup $\mathbb{L} = S\left( U(n_1) \times \cdots \times U(n_s) \right)$ left unbroken by the defect. Given a Levi subgroup $\mathbb{L}$, the additional label $\sigma$ specifies the nonequivalent choices for how $\mathbb{L}$ can be embedded into $SU(N)$. A particular choice can be denoted by $\mathbb{L}_\sigma$. The monodromy defect labelled by $\sigma=1$ corresponds to the singularity
\be
\oint_{|z_2|=\epsilon} A = 2 \pi \vec{m} \ ,
\ee
where
\be
	\vec{m} = (\underbrace{m_1,\cdots,m_1}_{n_1},\underbrace{m_2,\cdots,m_2}_{n_2},\cdots,\underbrace{m_s\cdots,m_s}_{n_s}) \ ,
\ee
and $m_1,\ldots,m_s$ are identified with the FI parameters of the three-dimensional theory $T_\rho(\mathfrak{g})$. Note that $\vec{m}$ can be characterized by the property $(\vec{m},\rho_{\mathfrak{l}}) = 0$ where $\rho_{\mathfrak{l}} = \rho_{n_1} \oplus \ldots \oplus \rho_{n_s}$ is the Weyl vector of the subalgebra $\mathfrak{l}$ with the embedding $\sigma=1$. The remaining supersymmetric vacua $\sigma$ correspond to nonequivalent choices of Levi subalgebra of the same Levi type $\mathfrak{l}$. Thus $\sigma$ correspond to permutations that are not simply permutations within each block. The number of such permutations is clearly $N_\rho = N! / ( n_1! \ldots n_s! )$ which matches the number obtained from the quiver description of $T_\rho$. Thus, the supersymmetric vacua are elements $\sigma \in \cW / \cW_{\mathfrak{l}}$ where $\cW_{\mathfrak{l}}$ is the Weyl group of $\mathfrak{l}$.

Let us now explain how to compute the 5d Nekrasov partition function in the presence of a monodromy defect using the ramified instantons computations of~\cite{Alday:2010vg,Kanno:2011fw}.
It is known that the moduli space of the ramified instantons of $U(N)$ gauge theory is equivalently described by the moduli space of the instantons on the orbifold space $\mathbb{C}\times \mathbb{C}/\mathbb{Z}_s$ where $\mathbb{Z}_s$ acts on the complex coordinates as $(z,w) \rightarrow (z,\omega w)$ with $\omega=e^{\frac{2\pi i}{s}}$. Note that the defect spans the $z$-plane. The equivalence between the monodromy defect and orbifold construction has been proven in the mathematical literature~\cite{2008arXiv0812.4656F,2010arXiv1009.0676F}. Reference~\cite{Bullimore:2014awa} (see also~\cite{Nawata:2014nca}) also demonstrates explicitly that the obrbifolding procedure directly reproduces the vortex partition function of $T_\rho(\mathfrak{g})$, when the 5d gauge coupling is sent to zero.

The geometric orbifold action is accompanied by the non-trivial $U(1)^s$ holonomy action on the gauge group which will be explained momentarily. We also twist the rotation symmetry $\mathfrak{so}(2)_1$ of the coordinate $w$ with the $\mathfrak{so}(2)_2$ R-symmetry subgroup in $\mathfrak{so}(5)_R$ to have unbroken supersymmetries. The $\mathbb{Z}_s$ then acts on the diagonal combination $\mathfrak{u}(1)_d$.

\begin{figure}[tb]
\begin{center}
\includegraphics[width=80mm]{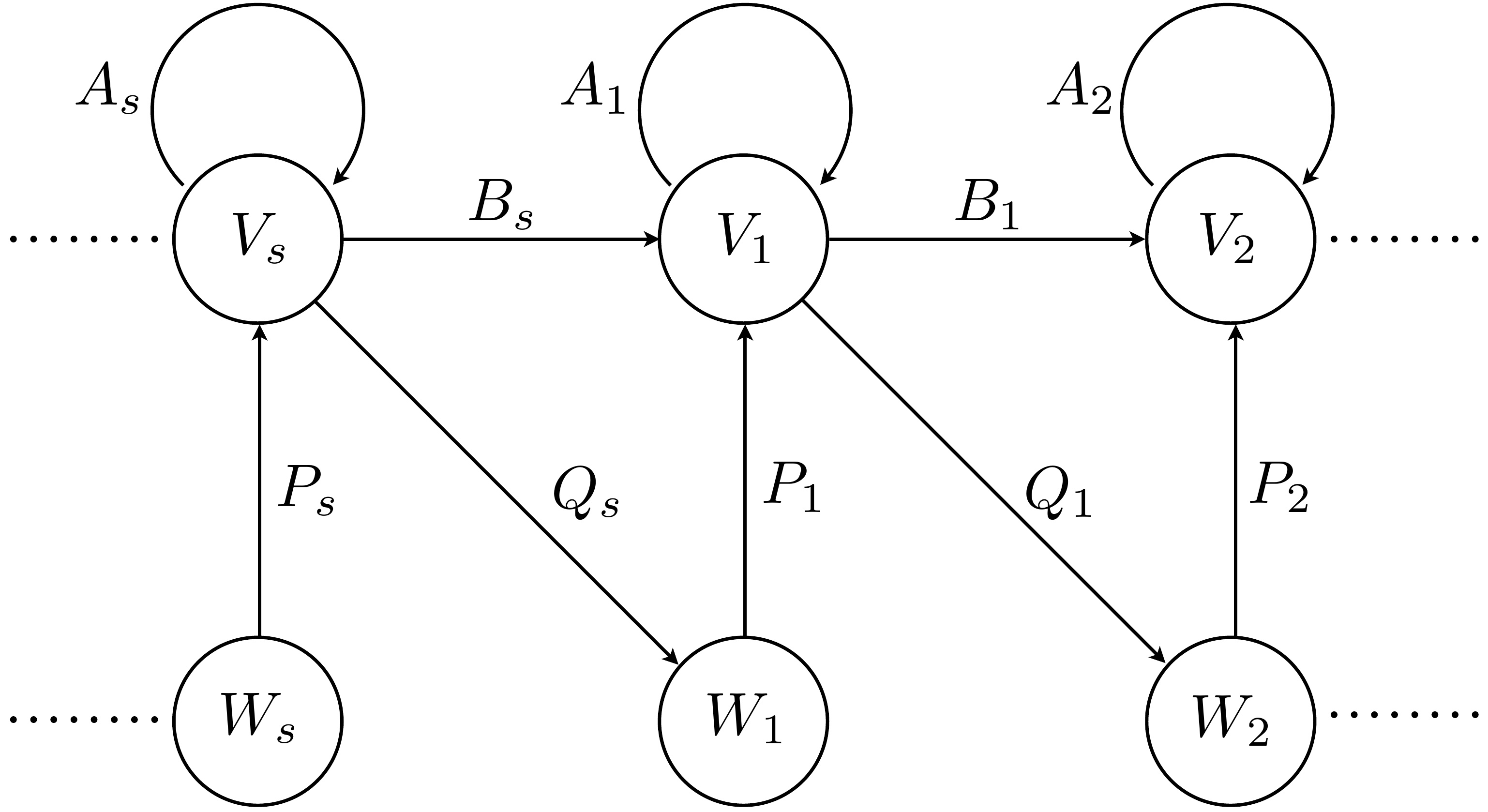}
\end{center}
\vspace{-20pt}
\caption{Chain-saw quiver}
\label{fig:chainsaw}
\end{figure} 

This allows us to construct the instanton moduli space with a monodromy defect from the usual ADHM construction simply by applying the $\mathbb{Z}_s$ action. The standard orbifolding procedure leads to an ADHM construction whose quiver, called the chain-saw quiver, is shown in Figure \ref{fig:chainsaw}. After the orbifolding, we have vector spaces $V_i$ and $W_i$ of complex dimensions
\be
    {\rm dim}_{\mathbb{C}} W_i = n_i \,, \quad {\rm dim}_{\mathbb{C}} V_i = k_i \ ,
\ee
for the nodes in the quiver diagram. Here the index $i$ is defined modulo $s$ so that $V_{s+1} = V_1$ and $W_{s+1}=W_1$. The associated ADHM data are given by matrices $A_i \in {\rm Hom}(V_i,V_i),\, B_i \in {\rm Hom} (V_i,V_{i+1}), \, P_i \in {\rm Hom} (W_i,V_i)$ and $Q_i \in {\rm Hom} (V_i,W_{i+1})$. As a complex manifold, the moduli space of the ramified instantons $\cM_{\rho,k_1,\cdots,k_s}$ is obtained by setting to zero the complex moment map
\be
    A_{i+1}B_i - B_i A_i + P_{i+1}Q_i = 0 \ ,
\ee
and performing a quotient by the complexified gauge group $\otimes_i GL(k_i,\mathbb{C})$.

The localization of the ramified instanton partition function was explained in~\cite{Alday:2010vg,Wyllard:2010vi,Kanno:2011fw}. The saddle points of the localization are again classified by the standard $N$-tuple of Young tableaux $\vec{Y}$.
The Young tableaux are now labelled by
\be
\vec{Y} = \{Y_{j,\alpha}\} \,, \quad (j=1,\cdots,s \,, \ \alpha = 1,\cdots,n_s) \ .
\ee
The boxes in the $j$-th column of the tableau $Y_{i,\alpha}$ contributes to the dimension of $V_{i+j-1}$, i.e. $k_{i+j-1}$.

Let us first compute the equivariant indices of vector bundles on the ramified instanton moduli space. We compute it from the equivariant index on the standard instanton moduli space by acting $\mathbb{Z}_s$ orbifold. The $\mathbb{Z}_s$ orbifold can be realized as an action on the equivariant parameters $\epsilon_2,m$ and $a$. Before proceeding, the parameters $a$ should be renamed as
\be
    (a_1,a_2,\cdots,a_N) = \{a_{i,\alpha}\} \ .
\ee
We turn on the $U(1)^s$ gauge holonomy such that $\mathbb{Z}_s$ action rotates the guage parameters as $a_{j,\alpha} \rightarrow a_{j,\alpha} - j\frac{2\pi ir}{s}$, while changing the $\mathfrak{so}(2)_1$ rotation parameter as $\epsilon_2\rightarrow \frac{\epsilon_2}{s} +\frac{2\pi i r}{s}$.
Alternatively, we can redefine the gauge parameters as
\be
    a_{j,\alpha} \rightarrow a_{j,\alpha} - j\epsilon_2 \ ,
\ee
and turn off the $U(1)^s$ holonomy, which effectively substitutes the $U(1)^s$ holonomy action by $\mathbb{Z}_s$ action only on $\epsilon_2$ parameter.

For the tangent bundle $\mathcal{TM}^{\rho}$ of the ramified instanton moduli space, the equivariant index at the fixed point $\vec{Y}$ is given by
\be
    {\rm Ind}_{\mathcal{TM}_{\vec{Y}}^{\rho}}(\epsilon_1,\epsilon_2,a) = \sum_{r=0}^{s-1}\frac{1}{s} {\rm Ind}_{\mathcal{TM}_{\vec{Y}}}(\epsilon_1,\frac{\epsilon_2}{s},a_{j,\alpha}-\frac{j\epsilon_2}{s}) \Big|_{\epsilon_2 \rightarrow \epsilon_2 + 2\pi ir} \ ,
\ee
and, for the adjoint hypermultiplet, we get the index
\be
    {\rm Ind}_{\mathcal{V}_{\vec{Y}}^{H,\rho}}(\epsilon_1,\epsilon_2,a,m) = \sum_{r=0}^{s-1}\frac{1}{s} {\rm Ind}_{\mathcal{V}^H_{\vec{Y}}}(\epsilon_1,\frac{\epsilon_2}{s},a_{j,\alpha}-\frac{j\epsilon_2}{s},m-\frac{\epsilon_2}{2s}) \Big|_{\epsilon_2 \rightarrow \epsilon_2 + 2\pi ir} \ .
\ee
Remember that the $\mathbb{Z}_s$ orbifold acts on the $\mathfrak{u}(1)_d$ which simultaneously rotates the coordinate $w$ and the $\mathfrak{so}(2)_2$ R-symmetry, and also on the Cartans of $\mathfrak{u}(N)$. In the above indices we have implemented the $\mathbb{Z}_s$ orbifold as the action only on $\epsilon_2$ by shifting the mass parameters $a$ and $m$. In addition, we need to multiply the momentum factor $\sum_{t\in \mathbb{Z}}e^{\frac{2\pi}{r} t}$ along the temporal circle in five-dimensions.

The partition function computation is straightforward using the conversion rule.
The ramified instanton partition function is given by
\be
    Z_{\rm inst}^{\rho}(\epsilon_1,\epsilon_2,a_{s,I},m,\mathfrak{q}) = \sum_{\vec{Y}} z(\vec{Y}) \prod_{i=1}^s \mathfrak{q}_i^{k_i(\vec{Y})} \ .
\ee
$z(\vec{Y})$ is the product of weights in the equivariant indices at the saddle point $\vec{Y}$ and $\mathfrak{q}_i$ are the instanton fugacities. $\mathfrak{q} \equiv \mathfrak{q}_1\mathfrak{q}_2\cdots \mathfrak{q}_s$ is related to the dynamical coupling of the bulk gauge theory. We identify the instanton fugacities with the monodromy parameters as follows:
\be
    \mathfrak{q}_{i=1,2,\cdots,s-1} = e^{\frac{4\pi^2 r}{g^2}(m_i-m_{i+1})} \,,\quad \mathfrak{q}_s = \mathfrak{q} e^{\frac{4\pi^2 r}{g^2}(m_s-m_1)} \ .
\ee

Similarly, we can compute the perturbative contribution under the $Z_s$ orbifold using the above equivariant indices. It is given by
\bea\label{eq:ramified-pert}
    Z_{\rm pert}^\rho \!&=\! \prod_{i,j=1}^s\!\prod_{\alpha=1}^{n_i}\prod_{\beta=1}^{n_j} 
    \left[ \frac{ \tilde\Gamma_3\big( a_{i,\alpha}\!-\!a_{j,\beta}\!-\! \lceil\frac{i\!-\!j}{s}\rceil\epsilon_2\!+\! m\!+\!\frac{\epsilon_1}{2} \big) \tilde\Gamma_3\big( a_{i,\alpha}\!-\!a_{j,\beta}\!-\! \lceil\frac{i\!-\!j\!-\!1}{s}\rceil\epsilon_2\!-\! m\!+\!\frac{\epsilon_1}{2} \big)  }
     { \tilde\Gamma_3'\big(a_{i,\alpha}\!-\!a_{j,\beta}\!-\! \lceil\frac{i\!-\!j}{s}\rceil\epsilon_2\big) \tilde\Gamma_3\big(a_{i,\alpha}\!-\!a_{j,\beta}\!-\! \lceil\frac{i\!-\!j\!-\!1}{s}\rceil\epsilon_2+\epsilon_1\big)} \right]^{1/2} \ ,
\eea
where $\lceil \cdot \rceil$ denotes the ceiling function.
It turns out that the perturbative partition function factorizes into the contributions from the 5d theory and from the 3d theory supported on the defect. For instance, for the full defect of type $\rho = [1,1,\cdots,1]$, the perturbative partition function factorizes as
\begin{align}
    Z_{\rm pert}^{\rho=[1^N]} &= Z_{\rm pert}(m\rightarrow m+\epsilon_2/2) \times Z_{\rm 3d,pert} \nonumber \ , \\
    Z_{\rm 3d,pert} &= \prod_{e>0} \left[\frac{ S_2'\left((e,a);\frac{2\pi}{r},\epsilon_1\right)S_2\left((e,a)-\frac{2\pi}{r};-\frac{2\pi}{r},\epsilon_1\right) }
    { S_2\left((e,a)+m+\frac{\epsilon_1}{2};\frac{2\pi}{r},\epsilon_1\right)S_2\left((e,a)+m-\frac{2\pi}{r}+\frac{\epsilon_1}{2};-\frac{2\pi}{r},\epsilon_1\right) }
     \right]^{1/2} \ ,
\end{align}
with the double sine function defined as the following regularized infinite product
\be
    S_2(z;w_1,w_2) \sim \prod_{n_1,n_2=0}^\infty\frac{(n_1w_1+n_2w_2+z)}{((n_1+1)w_1+(n_2+1)w_2-z )}
\ee
The primed function is defined as $S'_2(0) \equiv \lim_{z\rightarrow 0}S_2(z)/z$.

$Z_{\rm 3d,pert}$ is the perturbative contribution from the 3d theory on the defect. 
Indeed, this 3d factor agrees with the perturbative contribution in the holomorphic block of the three dimensional $T[U(N)]$ theory, which is believed to be the 3d theory living on the defect.
We find that the product of two 3d factors, from $S^3_{(3)}$ for example, with the physical parameters $\mu,\omega_{1,2,3}$ constructs the 1-loop contribution to the $S^3$ partition function of the $T[U(N)]$ theory in the Higgs branch expression~\cite{Bullimore:2014awa}:
\bea
    &Z_{\rm 3d,pert}(m=\mu+\omega_1/2+\omega_3/2,r=2\pi/\omega_1,\epsilon_1=\omega_2) \times Z_{\rm 3d,pert}(\omega_1 \leftrightarrow \omega_2) \\ 
    =& \prod_{e>0}\frac{ S_2((e,a);\omega_1,\omega_2) }
    { S_2((e,a)+\mu+\frac{\omega_1+\omega_2+\omega_3}{2};\omega_1,\omega_2) }  \ .
\eea
 Furthermore we find that the ramified instanton partition function in the decoupling limit $\mathfrak{q}\rightarrow 0$ (or $g\rightarrow 0$) reproduces the vortex partition function of the $T[U(N)]$ theory.
For example, the instanton partition function of the $U(2)$ gauge theory in the decoupling limit is
\bea
\lim_{\mathfrak{q}\rightarrow0}Z^{\rho=[1^2]}_{\rm inst} =& 1 + \frac{\sin\frac{r(-m+\epsilon_1/2)}{2}\sin\frac{r(a_{12}+m-\epsilon_1/2)}{2}}{\sin\frac{r\epsilon_1}{2}\sin\frac{r(a_{12}-\epsilon_1)}{2}}\mathfrak{q}_1
 \\
 &+\frac{\sin\frac{r(-m+\epsilon_1/2)}{2}\sin\frac{r(a_{12}+m-\epsilon_1/2)}{2}\sin\frac{r(-m+3\epsilon_1/2)}{2}\sin\frac{r(a_{12}+m-3\epsilon_1/2)}{2}}{\sin\frac{r\epsilon_1}{2}\sin r\epsilon_1 \sin\frac{r(a_{12}-\epsilon_1)}{2}\sin\frac{r(a_{12}-2\epsilon_1)}{2}}\mathfrak{q}_1^2 + \cdots
\eea
This is precisely the vortex partition function of the $T[U(2)]$ theory~\cite{Pasquetti:2011fj,Bullimore:2014awa}.

Let us now discuss the limits $m = \epsilon_\pm $ which are needed for the chiral algebra limit of the superconformal index. Due to the mass shift $m\rightarrow m-\frac{\epsilon_2}{2}$, these limits become the limits $m=\pm\epsilon_1/2$. For simplicity we shall consider a particular embedding $\sigma=1$ of the Levi subgroup.

In the limit $m=-\epsilon_1/2$, the perturbative contribution (\ref{eq:ramified-pert}) reduces to unity due to the cancellation between contributions from vector and hypermultiplets
\be
    \lim_{m\rightarrow \epsilon_1/2} Z_{\rm pert}^{\rho} = 1 \ .
\ee
A similar cancellation happens in the instanton calculus and we find that the contribution at each instanton fixed point becomes unity. Thus the instanton series is significantly simplified so that it simply counts the number of fixed points characterized by the same instanton numbers $(k_1,k_2,\cdots,k_s)$.
For generic $\rho$, we claim that
\bea\label{eq:defect-instanton-limit}
    \lim_{m\rightarrow \epsilon_1/2} Z_{\rm inst}^\rho &= \left(\mathfrak{q};\mathfrak{q}\right)_\infty^{-N} \prod_{i=1}^s\prod_{j=1}^\infty\left(1-\prod_{a=i}^{i+j-1}\mathfrak{q}_a\right)^{-n_i} \nonumber \\
    &= \left(\mathfrak{q};\mathfrak{q}\right)_\infty^{-N} \prod_{i=1}^s\prod_{j=i+1}^s\left(e^{\frac{4\pi^2r}{g^2}m_{ij}};\mathfrak{q}\right)_{\!\infty}^{-n_i}\prod_{i=2}^s\prod_{j=1}^{i-1}\left(\mathfrak{q}e^{\frac{4\pi^2r}{g^2}m_{ij}};\mathfrak{q}\right)_{\!\infty}^{-n_i} \ .
\eea
where the $q$-Pochhammer symbol is defined as $(x;q)_\infty=\prod_{i=0}^\infty(1-xq^i)$ and we used the notation $m_{ij} = m_i-m_j$. 
The first equality was given in~\cite{Kanno:2011fw}. Here the index $a$ is taken to be modulo $s$. We have checked the second equality for $N=2,3,4,5,6,7$ with arbitrary $\rho$ at some lower orders in $\mathfrak{q}_i$ expansions.

In the second limit $m=\epsilon_1/2$, after some algebra and using identities of $S_2$, we find that the perturbative contribution simplifies to
\be
    \lim_{m\rightarrow \epsilon_1/2} Z_{\rm pert}^{\rho} = \left(\frac{r}{2\pi}\right)^{N/2}\prod_{i=1}^s\prod_{e\in\Delta_i^+} 2\sin(\frac{r}{2}(e,a)) \,,
\ee
where $\Delta^+_i$ denotes the positive roots of subgroup $U(n_i)\subset \mathbb{L}$.
For instantons, the contribution at each fixed point $\vec{Y}$ contains the following center of mass factor as a universal prefactor
\be
    \frac{\sin\frac{r(m-\epsilon_1/2)}{2}}{\sin\frac{r\epsilon_1}{2}} \,,
\ee
which vanishes when $m=\epsilon_1/2$. Therefore all fixed point contributions are identically 0 and it proves that
\be
    \lim_{m\rightarrow -\epsilon_1/2} Z_{\rm inst}^\rho = 1 \,.
\ee


\section{$W_N$ - Algebra Characters}\label{sec:walgebra}

We will now combine the results of the previous section to compute of the chiral algebra limit of the 6d superconformal index in the presence of supersymmetry preserving configurations of defects. In this limit, we can evaluate the Coulomb branch integral of the $S^5$ partition function explicitly and express the result manifestly as a 6d superconformal index.

\begin{figure}[tb]
\centering
\includegraphics[width=9cm]{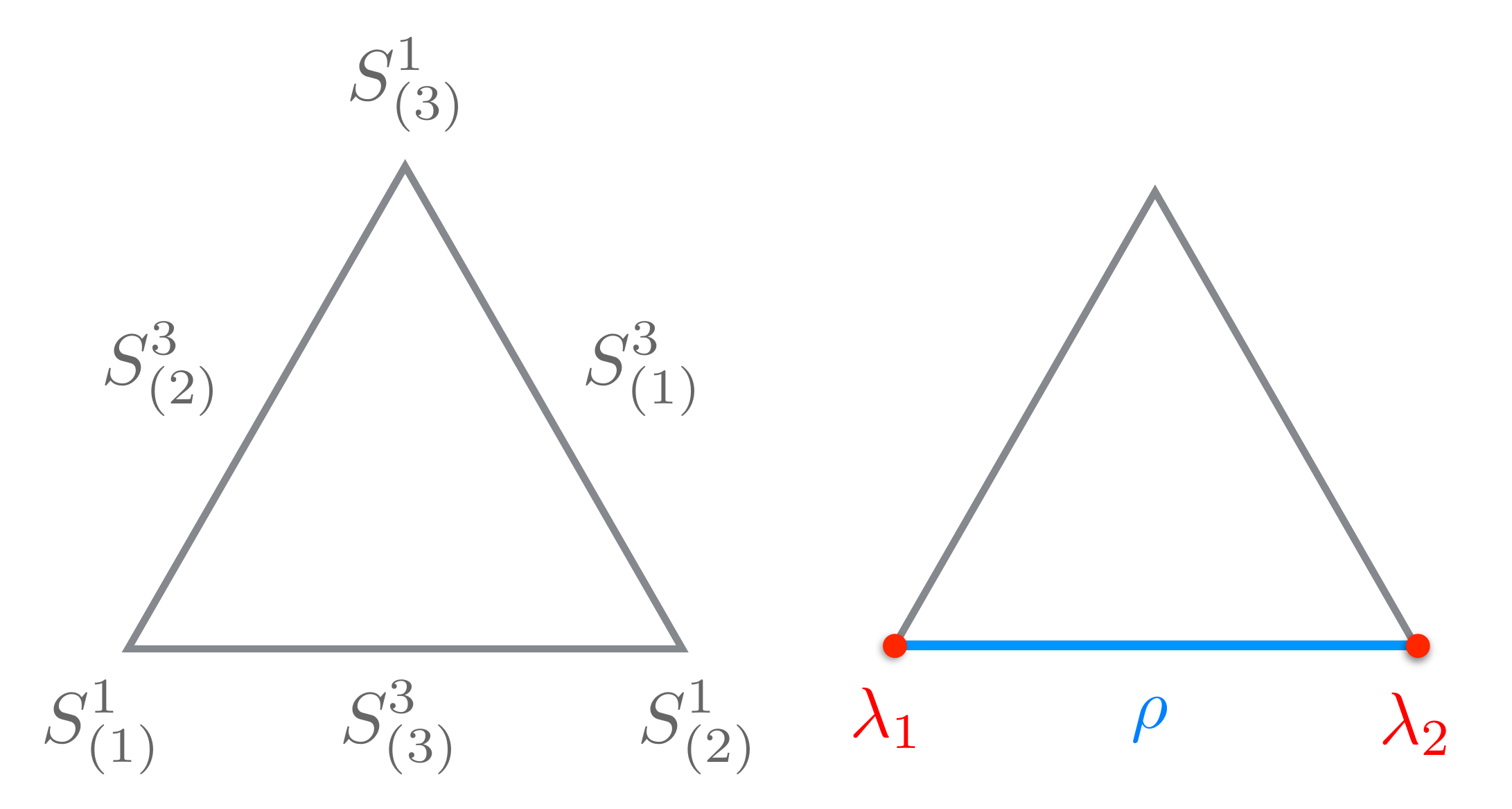}
\caption{\emph{A summary of the configurations of a codimension 2 and codimension 4 defects that in the chiral algebra limit reproduce characters of irreducible modules of the $W_N$ - algebra.}}
\label{fig:walgebra}
\end{figure}

In the absence of defects this superconformal index has been shown to coincide with the vacuum character of $\mathfrak{u}(1)$ in the case of the abelian tensormultiplet and the $W_N$ algebra for the non-abelian theory of type $A_{N-1}$~\cite{Kim:2012ava,Kim:2012qf,Beem:2014kka}. In this section, we consider combinations defects that do not intersect the fixed circle $S^1_{(3)} \subset S^5$ that is distinguished by the chiral algebra limit - as shown in figure~\ref{fig:walgebra}. From the perspective of the superconformal index, this means that the defects are point-like in the chiral algebra plane and are expected to correspond to chiral vertex operators. Indeed, we will reproduce the characters of irreducible modules of $\mathfrak{u}(1)$ and the $W_N$ - algebras found in~\cite{Drukker:2010jp}.

\subsection{Vacuum Module}\label{subsec:wvac}

Let us first review the computation of the chiral algebra limit of the 6d superconformal index in the absence of any defects~\cite{Kim:2012ava,Kim:2012qf}.

Tuning the mass parameter in the 5d partition function to $\mu \to \frac{1}{2}(\omega_1+\omega_2-\omega_3)$, the computation of the five-sphere partition function simplifies dramatically. In particular, the instanton partition functions at fixed points (1) and (2) are one and at fixed point (3) becomes simply a $q$-Pochhammer symbol. In addition, the perturbative contributions at fixed point (3) become one and simplify dramatically at fixed points (1) and (2). Let us consider the abelian tensormultiplet and the non-abelian theories in turn.

It is convenient to introduce the notation $2\pi i \tau = -\beta \omega_3$ so that $q=e^{2\pi i \tau}$. We also set $\omega_1\omega_2=1$ since the final results do not depend on it (i.e. only depend on the ratio $b^2 \equiv \omega_1/\omega_2 $). For the $U(1)$ theory, the contributions from each fixed point are
\bea
Z_{(1)} = Z_{(2)} = 1 \,, \qquad Z_{(3)} = 1/\eta(-1/\tau) \, .
\eea
We have multiplied to $Z_{(3)}$ an overall factor $e^{\frac{\pi i}{12\tau}}$ to make it as a modular form. As studied in \cite{Kim:2012qf} this factor is related to the leading high temperature behavior of the 6d abelian index, which cannot be observed from the 5d partition function because we have assumed the index to be smooth in 5d limit and regularized it. The detailed discussion will be left for a later work~\cite{BBK2014}.
Combining these contributions with the classical contribution $e^{\frac{\pi i a^2}{\tau}}$ we have a gaussian integral
\be
Z_{U(1)} = i\int_{i \mathbb{R}} da \, \frac{ e^{\pi i a^2 / \tau }}{\eta(-1/\tau)} = \frac{1}{\eta(\tau)} \, .
\ee
This expression is the vacuum character of the $\widehat{\mathfrak{u}}(1)$ chiral algebra including the contribution from the central charge $c=1$.

For the $SU(N)$ theory the contributions from each fixed point are
\be
Z_{(1)} = \prod\limits_{e>0} 2 \sin \pi b^{-1}( e , a ) \,, \qquad 
Z_{(2)} = \prod\limits_{e>0} 2 \sin \pi b( e , a ) \,, \qquad
Z_{(3)} = 1 / \eta(-1 /\tau)^{N-1} \ .
\ee
Strictly speaking we have performed the instanton calculus for $U(N)$ and divided by the $U(1)$ instanton contribution. Combining these contributions with the classical contribution $e^{\frac{\pi i}{\tau} (a,a)}$ we find
\bea
Z_{SU(N)} & = \int [da] \, \prod_{e >0} 2 \sin \pi b^{\pm}( e, a ) \, \frac{ e^{ \frac{\pi i}{ \tau} (a,a)}}{\eta(-1/\tau)^{N-1}}  \\ 
& = \frac{q^{-\frac{1}{2}(Q,Q)}}{\eta(\tau)^{N-1}}  \sum_{\sigma\in \cW} \epsilon(\sigma) q^{-(\sigma(\rho)-\rho,\rho)} \ .
\eea
For compactness we have introduced the shorthand notation $f(b^{\pm}) =f(b)f(b^{-1})$. We have also introduced the standard notation $Q=(b+1/b) \rho$ where $\rho = \sum_{j=1}^N \omega_j $ is the Weyl vector and we have a summation over the Weyl group $\cW$ of $\mathfrak{g} = A_{N-1}$. The integration over $a$ in the Cartan subalgebra was again gaussian and performed by systematically completing the square in the exponential.
 
Now, using the Weyl denominator formula
\be
\sum_{\sigma\in \cW} \epsilon(\sigma) q^{-(\sigma(\rho)-\rho,\rho)}  = \prod_{e>0}(1-q^{(\rho,e)}) \ ,
\ee
we can express the partition function on squashed $S^5$ in a form that is manifestly a superconformal index or partition function on $S^1 \times S^5$,
\bea
\frac{q^{-\frac{1}{2}(Q,Q)}}{\eta(\tau)^{N-1}}  \sum_{\sigma\in \cW} \epsilon(\sigma) q^{-(\sigma(\rho)-\rho,\rho)} & = q^{-\frac{c}{24}} \, \frac{\prod\limits_{e>0}(1-q^{(\rho,e)})}{ \prod\limits_{n=1}^{\infty}(1-q^n)^{N-1} } \\
& = q^{-\frac{c}{24}} \mathrm{PE}\left[ \frac{(N-1)q}{1-q} - \sum_{e>0} q^{(\rho,e)}  \right] \\
& = q^{-\frac{c}{24}} \mathrm{PE}\left[ \frac{(N-1)q}{1-q} - \sum_{j=1}^{N-1} j q^{N-j}  \right] \\
& = q^{-\frac{c}{24}} \mathrm{PE}\left[ \frac{q^2+\ldots + q^N}{1-q} \right] \, .
\label{vacWN}
\eea
where
\bea
c & = N-1+12 (Q,Q) \\
& = (N-1) + N(N^2-1) (b+1/b)^2\, .
\eea

This result is precisely the vacuum character of the $W_N$ -  algebra, see appendix~\ref{asubsec:wn}. In particular, the final lines of equation~\eqref{vacWN} reflect that the states $W^{(n)}_{-l} | 0 \ra $ are null if $0 < l <n$ and that the vacuum module is freely generated by $W_{-l}^{(n)}$ for $n=2,\ldots,N$ and $l \geq n$. In our correspondence with the superconformal index of the $(2,0)$ theory, we can identify these generators with local operators $\cO_n$ for $n=1,\cdots,N-1$ generating the $\frac{1}{2}$-BPS chiral ring and their holomorphic derivatives i.e. $W_{-l}^{(n)}  = \del^{l-n} \cO_n$. Note that all the states contributing to the superconformal index are bosonic.


\subsection{Degenerate Modules}\label{subsec:wdeg}

In this subsection, we enrich the above computation by adding supersymmetric Wilson loops wrapping $S^1_{(1)}$ and $S^1_{(2)}$. We expect to find non-vacuum modules of the relevant chiral algebra. In the abelian tensormultiplet theory, we find non-vacuum modules whose dimension depends on a pair of integers $n_1$ and $n_2$. For the non-abelian theory of type $A_{N-1}$, we will find the characters of the so-called completely degenerate modules of the $W_N$ - algebra.

Let us first consider the abelian tensormultiplet theory and add supersymmetric Wilson loops of integer charge $n_1$ and $n_2$ on the circles $S^1_{(1)}$ and $S^1_{(2)}$ respectively. As described section~\ref{subsec:codim4}, the presence of the Wilson loops modifies the instanton partition functions localized at fixed points $(1)$ and $(2)$. However, in the special limit the instantons at these fixed points decouple and the contribution is simply the classical expectation values. In summary we have
\be 
Z_{(1)} = e^{2\pi i a n_1 /b } \,, \qquad Z_{(2)} = e^{2\pi i a n_2 b } \,, \qquad Z_{(3)} = 1 / \eta(-1/\tau) \, .
\ee
Combining with the classical contribution, we again have a gaussian integral
\bea
Z_{U(1)}^{(n_1,n_2)} & =i \int_{i \mathbb{R}} da \, e^{2\pi i a n_1 / b } e^{2\pi i a n_2 b } \, \frac{ e^{\pi i a^2 / \tau }}{\eta(-1/\tau)} \\
& = \frac{e^{-i\pi \tau(n_1 / b+n_2b)^2}}{\eta(\tau)} \ .
\eea
This expression is the character of an irreducible non-vacuum module of $\widehat{\mathfrak{u}}(1)$ with dimension $\Delta = - \frac{1}{2}(n_1/b+n_2 b)^2$. 

In the non-abelian case, we can add supersymmetric Wilson loops supported on the circles $S^1_{(1)}$ and $S^1_{(2)}$ and labelled by irreducible representations of $A_{N-1}$ with highest weights $\lb_1$ and $\lb_2$ respectively. Let us first consider the case where $\lb_2=0$ in some detail. As above, the instanton contributions at fixed points $(1)$ and $(2)$ decouple in the special limit and the contributions from the fixed points are
\be
Z_{(1)} = \prod\limits_{e>0} 2 \sin \pi b^{-1} ( e , a ) \, \tr_\lambda (e^{2\pi i a/b} )\,, \quad
\quad Z_{(2)} =  \prod\limits_{e>0} 2 \sin \pi b ( e , a ) \,, \quad
\quad Z_{(3)} = 1 / \eta(-1/\tau)^{N-1} \ ,
\ee
The classical value of the Wilson loop in the irreducible representation of highest weight $\lambda$ is inserted to $Z_{(1)}$. We have also multiplied a factor $e^{\frac{(N-1)\pi i}{12\tau}}$ to $Z_{(3)}$ by hand. Computing the gaussian integral in this case we find
\bea
Z_{\lb}(\tau) & =  \int [da] \, \prod_{e >0} 2 \sin \pi b^{\pm} ( e, a ) \, \frac{ e^{ \frac{\pi i}{ \tau } (a,a)}}{\eta(-1/\tau)^{N-1}} \tr_{\lb}(e^{2\pi i a/b })  \\ 
& = \frac{q^{\Delta(\mu)-\frac{1}{2}(Q,Q)}}{\eta(\tau)^{N-1}}  \sum_{\sigma\in \cW} \epsilon(\sigma) q^{-(\sigma(\rho)-\rho,\rho+\lambda)} \ ,
\eea
which is precisely the character of a completely degenerate representation of the $W_N$ - algebra with momentum $\mu = -\lambda/b$ and dimension $\Delta(\mu) = (Q,\mu)-\frac{1}{2}(\mu,\mu)$.

It is again illuminating to express this result in terms of the Plethystic exponential. Using the formula
\be
\sum_{\sigma\in \cW} \epsilon(\sigma) q^{-(\sigma(\rho)-\rho,\rho+\lb)}  = \prod_{e>0}(1-q^{(\rho+\lb,e)}) 
\ee
we find 
\be
Z_\lb(\tau) = q^{\Delta(\mu) - \frac{c}{24}} \; \PE{ \frac{(N-1)q}{1-q} - \sum_{e>0}q^{(\rho+\lb,e)} } 
\ee
This demonstrates that we have a null state at level $(\rho+\lb,e)$ for each positive root $e>0$. For instance, in the case $N=2$ we find the character of the degenerate module of the Virasoro algebra with a null vector at level $r$
\be
q^{\Delta(\mu)- \frac{c}{24} } \; \PE{\frac{q}{1-q}-q^r } \ ,
\ee
where now $c = 1+6(b+1/b)^2$.

The most general completely degenerate module is found by placing two codimension 4 defects labelled by $\lb_1$ and $\lb_2$ wrapping the circles $S^1_{(1)}$ and $S^1_{(2)}$ respectively. The partition function now evaluates to
\bea
Z_{\lb_1,\lb_2}(\tau) & = \frac{q^{\Delta(\al)-\frac{1}{2}(Q,Q)}}{\eta(\tau)^{N-1}}  \sum_{\sigma\in \cW} \epsilon(\sigma) q^{-(\sigma(\rho+\lb_2)-\rho-\lb_2,\rho+\lambda_1)} \\
\eea
corresponding to a simple module with momentum $\al = -\lb_1/b - \lb_2 b$. This exhausts the spectrum of fully degenerate modules.


\subsection{Semi-Degenerate Modules}\label{subsec:wsemi}

In this subsection, we consider the a surface defect supported on the three-sphere $S^{3}_{(3)}$. From the perspective of the superconformal index this corresponds to a codimension 2 defect orthogonal to the chiral algebra plane. 

Let us briefly consider the abelian tensormultiplet theory. As we have argued in section, the presence of a monodromy defect does not change the instanton contributions in this case. The only modification comes from a classical contribution $e^{-2\pi i m a}$ where $m$ is the monodromy parameter. The $S^5$ partition function
\be
Z_{U(1)} = i\int_{i \mathbb{R}} da \, e^{-2\pi i m a} \, \frac{ e^{\pi i a^2 / \tau }}{\eta(-1/\tau)} = \frac{q^{-m^2/2}}{\eta(\tau)} \, .
\ee
is nothing but the character of an irreducible module of $\widehat{\mathfrak{u}}(1)$ of dimension $\Delta=-m^2/2$. We remind the reader that $m$ is imaginary in our notation.

In the non-abelian case, let us first consider the most straightforward codimension 2 defect of type $\rho=[1,1,\ldots,	1] = \varnothing$. This leaves unbroken the subgroup $\mathbb{L}_{\varnothing} = S\left( U(1) \times \ldots \times U(1) \right)$. There are $N!$ supersymmetric vacua labelled by an element $\sigma \in \cW$ and corresponding to a permutation of the monodromy parameters $\vec{m} = (m_1,\ldots,m_N)$. In what follows, we denote the monodromy parameters $\vec{m}$ by simply $m$ (or $\mu=Q+\vec{m}$), which should not be confused with the $\mathcal{N}=1^*$ mass parameter that we have already tuned to the special value.

In the chiral algebra limit, we find that the contributions from the fixed points are independent of the permutation $\sigma$. The contributions from the fixed circles $S^1_{(1)}$ and $S^1_{(2)}$ are one: $Z^{\varnothing,\sigma}_{(1)} = Z^{\varnothing,\sigma}_{(2)} = 1$. The contributions from the fixed circle $S^1_{(3)}$ is $Z^{\varnothing,\sigma}_{(3)} = 1 / \eta(-1/\tau)^{N-1}$. There are now two classical contributions in the presence of a codimension 2 defect. In addition to the familiar classical contribution $e^{ \frac{\pi i}{ \tau}(a,a)}$ there is also a contribution $e^{ - 2 \pi i (\sigma(m),a)}$ which depends on the supersymmetric vacuum $\sigma$. The origin of the factor is explained in the appendix~\ref{sec-S5-partitionftn}. 

Putting the contributions together and summing over the permutations we have only to perform a gaussian integral
\bea
\int [da] \, \sum_{\sigma \in S_N} e^{ -2 \pi i  (\sigma(m),a) }  \, \frac{ e^{ \frac{\pi i}{ \tau }(a,a)}}{\eta(-\frac{1}{\tau})^{N-1}} & = \frac{e^{-\pi i \tau (m,m)}}{\eta(\tau)^{N-1}} \\
& = q^{\Delta(\mu)-\frac{c}{24}}\, \mathrm{PE}\left[ \frac{(N-1)q}{1-q}\right] \, .
\eea
where we have defined $\mu=Q+m$ and as before $\Delta(\mu)= (\mu,Q)-\frac{1}{2}(\mu,\mu)$. This expression is precisely the character $\tr_{V_\mu} \left(  q^{L_0 - c /24} \right)$ of a non-degenerate irreducible module with momentum $\mu = Q+m$.

The above computation can be rephrased in an interesting way. First note that the classical and instanton contributions to the integrand combine to form the non-degenerate character $\tr_{V_{\al}}\left( \tilde{q}^{L_0 - c/24} \right)$ on a torus with complex structure $\tilde \tau = -1/\tau$ and with momentum $\al = Q+a$. The above integral is then
\be
\tr_{V_{\mu}}q^{L_0 - c/24}  = \int [da] \, \mu(a)  \, Z(m,a)  \, \tr_{V_{\al}}\tilde{q}^{L_0 - c/24} \ ,
\ee
where $\mu(a) = \prod\limits_{e > 0} 2\sin \pi b^\pm (e,a)$ is the 5d $\cN=1$ vectormultiplet measure and
\be
Z(m,a) = \frac{\sum_{w \in S_N} e^{- 2 \pi i (w(m),a) } }{\prod_{e > 0} 2\sin \pi b^\pm (e,a)} \, .
\ee
This is precisely the squashed $S^3$ partition function of the 3d $\cN=4$ theory $T(U(N))$ in the chiral algebra limit. This is an important evidence that our computation in terms of monodromy defects is correctly reproducing the surface defect that we intended. Note that the combination $S_{\mu,\al} = \mu(a) Z(a,m)$ can be identified with the modular transformation matrix for non-degenerate $W_N$ - characters.

Let us now consider a generic codimension 2 defect labelled by the partition $\rho = [n_1,n_2,\cdots,n_s]$ where $n_1 \leq n_2 \leq \cdots \leq n_s$ and $\sum_{i=1}^s n_i = N$. In the presence of the defect, the gauge symmetry of the five-dimensional theory is broken to the Levi type $\mathfrak{l}=\mathfrak{s}[\mathfrak{u}(n_1) \times \cdots \times \mathfrak{u}(n_s)]$. The supersymmetric vacua are labelled by a permutation $\sigma \in \cW / \cW_{\mathfrak{l}}$. Due to the presence of non-abelian factors in $\mathfrak{l}$, there are now non-trivial perturbative contributions
\be
Z^{\rho,\sigma}_{(1)} = \prod_{j=1}^s \prod_{e \in \Delta_{j}^+} 2\sin \pi b^{-1} (e,\sigma(a)) \,, \qquad Z^{\rho,\sigma}_{(2)} = \prod_{j=1}^s \prod_{e \in \Delta_{j}^+} 2\sin \pi b (e,\sigma(a)) \ ,
\ee
where $\Delta_j^+$ corresponds to the positive root space generated by $\{ e_{r_j} , \ldots, e_{r_{j+1}-1} \}$ where $r_j = n_1+\ldots+n_j$. These are the roots whose non-zero elements lie entirely within the $n_j \times n_j$ block. The instanton contributions to the the fixed point $S^1_{(3)}$ remain unaffected by the presence of the defect, so $Z_{(3)} = 1/\eta(-1/\tau)^{N-1}$. As before, there is an additional classical contribution $e^{-2\pi i(\sigma(m),a)}$ - see appendix~\ref{sec-S5-partitionftn}.

Putting the contributions together and summing over the supersymmetric vacua we have
\bea
	&\int [da] \sum_{\sigma \in \cW / \cW_{\mathfrak{l}} }
	\, Z^{\rho,\sigma}_{(1)} Z^{\rho,\sigma}_{(2)} \,
	 e^{ -2 \pi i (\sigma(m),a) }  
	\, \frac{ e^{ \frac{\pi i}{ \tau }(a,a)}}{\eta(-\frac{1}{\tau})^{N-1}}  \\
	=& \frac{q^{\Delta(\mu) -\frac{1}{2}(Q,Q)}}{\eta(\tau)^{N-1}}
	\sum_{\sigma \in \mathcal{W}_{\mathfrak{l}}}\epsilon(\sigma)  q^{-( \sigma(\rho_{\mathfrak{l}}) - \rho_{\mathfrak{l}} ,\rho_{\mathfrak{l}})} \ ,
\eea
which is the character $\tr_{V_\mu} \left( q^{L_0- c/24} \right)$  of a semi-degenerate module of the $W_N$ - algebra with momentum $\mu = Q + m-(b+b^{-1})\rho_\mathfrak{l}$ where $(m,\rho_\mathfrak{l})=0$ and $\rho_\mathfrak{l}$ is the Weyl vector of the subalgebra $\mathfrak{l}$. 

We can also introduce the codimension 4 defects supported on the fixed circles $S^1_{(1)}$ and $S^1_{(2)}$. As these circles are inside $S^3_{(3)}$ we may only introduce supersymmetric Wilson loops in the unbroken gauge symmetry. For the abelian theory, we can introduce supersymmetric Wilson loops of any integer charges $n_1$ and $n_2$, and the result is the character of an irreducible module of $\widehat{\mathfrak{u}}(1)$ of dimension $\Delta = -(m-n_1/b-n_2 b)^2/2$.

For the non-abelian theories, the supersymmetric Wilson loops at the fixed circles $S^1_{(1)}$ and $S^1_{(2)}$ are characterized by the dominant integral weights $\lambda_1$ and $\lambda_2$ of the the Levi subalgebra $\mathfrak{l} \subset \mathfrak{g}$, respectively. The weights obey the constraints $(\lambda_1,e)\ge0$ and $(\lambda_2,e)\ge0$ for all $e \in \bigcup_{j=1}^s \Delta^+_j$. Plugging these Wilson loop contributions into the partition function we obtain
\bea
	&\int [da]
	\, \sum_{\sigma \in \cW / \cW_{\mathfrak{l}}} \, Z^{\rho,\sigma}_{(1)} Z^{\rho,\sigma}_{(2)}  \,
	\tr_{\lambda_1} (e^{2\pi i a/b} )  \, \tr_{\lambda_2} (e^{2\pi i a b } ) \,
	 e^{ -2 \pi i (\sigma(m),a) }  
	\, \frac{ e^{ \frac{\pi i}{ \tau }(a,a)}}{\eta(-\frac{1}{\tau})^{N-1}}  \\
	=& \frac{q^{\Delta(\mu)-\frac{1}{2}(Q,Q)}}{\eta(\tau)^{N-1}}
	\sum_{\sigma\in \mathcal{W}_{\mathfrak{l}}}\epsilon(\sigma) q^{(\rho_{\mathfrak{l}}+\lambda_1,\rho_{\mathfrak{l}}+\lambda_2)-(\sigma(\rho_{\mathfrak{l}}+\lambda_1),\rho_{\mathfrak{l}}+\lambda_2)} \ ,
\eea
where $\mu = Q + m -b^{-1}(\rho_\mathfrak{l}+\lb_1) - b(\rho_\mathfrak{l}+\lb_2)$. Here $\mathcal{W}_{\mathfrak{l}}$ is the Weyl group and $\rho_{\mathfrak{l}}$ is the Weyl vector of $\mathfrak{l}$.


\section{Affine Characters}\label{sec:affine}

In this section, we will consider adding a surface defect wrapping one of the circles $S^3_{(1)}$ or $S^3_{(2)}$. From the perspective of the 6d superconformal index this corresponds to adding a codimension 2 defect wrapping the chiral algebra plane. In the case of a codimension 2 defect labelled by the partition $\rho=[1^N]$ we will now find characters of irreducible modules of the affine algebra $\widehat{\mathfrak{su}}(N)$ at level $k = - N - b^{\pm2}$. We will leave exploration of general type $\rho$ defects for the future.

\begin{figure}[tb]
\centering
\includegraphics[width=13cm]{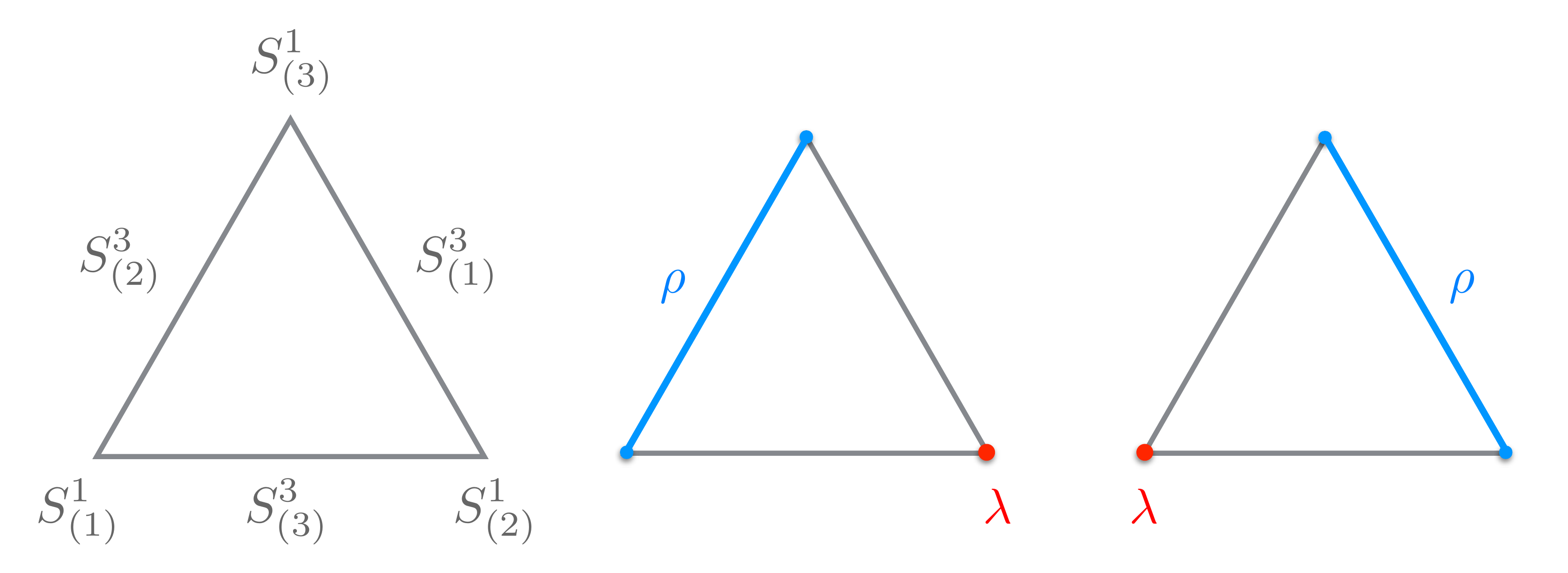}
\caption{\emph{A summary of the configurations of a codimension 2 and codimension 4 defects that in the chiral algebra limit reproduce characters highest weight modules of the affine $\widehat{\mathfrak{su}}(N)$ at level $k = - N - b^{\pm2}$.}}
\label{fig:affine}
\end{figure}


\subsection{Vacuum Module}\label{subsec:avac}
Let us first consider a surface defect of type $\rho$ supported on $S^3_{(1)}$. The same formulae will apply for $S^3_{(2)}$ by interchanging $b \leftrightarrow 1/b$. These correspond to codimension 2 defects wrapping the chiral algebra plane. In the chiral algebra limit, the contributions localized at the first two fixed points are
\be
Z_{(1)} = \prod_{e>0} 2 \sin \pi b^{-1} (e,a) \,,\qquad
\qquad Z^{\rho,\sigma}_{(2)} = \prod_{j=1}^s \prod_{e \in \Delta_{j}^+} 2\sin \pi b (e,\sigma(a)) \ ,
\ee
and, from (\ref{eq:defect-instanton-limit}), at the third fixed point is
\be
    Z_{(3)}^{\rho} = \left(\tilde{q};\tilde{q}\right)_\infty^{-N} \prod_{i=1}^s\prod_{j=i+1}^s\left(e^{\frac{2\pi i}{\tau}m_{ij}};\tilde{q}\right)_{\!\infty}^{-n_i}\prod_{i=2}^s\prod_{j=1}^{i-1}\left(\tilde{q}e^{\frac{2\pi i}{\tau}m_{ij}};\tilde{q}\right)_{\!\infty}^{-n_i} \ .
\ee
Combining all three fixed point contributions, the full partition function is
\begin{align}
	\,& \int [da] \prod_{e>0} 2 \sin \pi b^{-1} (e,a) \sum_{\sigma\in \mathcal{W}/\mathcal{W}_{\mathfrak{l}}}  \epsilon(\sigma)  
	\prod_{j=1}^s \prod_{e \in \Delta_{j}^+} 2\sin \pi b (e,\sigma(a)) \ e^{\frac{\pi i}{\tau}(a,a)-\frac{2\pi i}{\tau}b (\sigma(m),a)}  \times Z_{(3)} ^{\rho} \nonumber \\
	=& (-i\tau)^{\frac{N-1}{2}}q^{\Delta(\mu)-\frac{1}{2}(Q,Q)} \sum_{\sigma\in \mathcal{W}} \epsilon(\sigma) q^{(\sigma(m-\rho_{\mathfrak{l}})-m+\rho_{\mathfrak{l}},\rho)} \times  Z_{(3)}^{\rho} \ ,
\end{align}
where $\mu = Q + m - b^{-1}\rho - b\rho_{\mathfrak{l}}$.
This is not yet the superconformal index as the function $Z_{(3)}^\rho$ still takes the form of a weak-coupling expansion expanded by $\tilde{q}$ not $q$. The superconformal index can be in principle obtained by S-dualizing $Z_{(3)}^\rho$ and putting the result to the rest of the partition function. However, for generic $\rho$, the S-dual formulae for $Z_{(3)}^\rho$ is not known and we could not find the index expression of the partition function. 

In what follows we shall focus on a surface defect of maximal type $\rho=[1^N]$ in the chiral algebra limit for which $Z_{(3)}$ has a nice modular property and can be re-expanded by $q$. We find
\bea
    Z_{(3)}^{\rho=[1^N]}
    &= \eta\left(-1/\tau\right)^{\frac{(N-1)(N-2)}{2}} \prod_{e<0} \theta\left( \frac{(e,m)}{\tau} \Big| -\frac{1}{\tau} \right)^{-1} \\
    &= (-i)^{-\frac{N(N-1)}{2}}(-i\tau)^{-\frac{N-1}{2}}  \eta(\tau)^{\frac{(N-1)(N-2)}{2}} \prod_{e<0} e^{-\frac{\pi i (e,m)^2 }{\tau}} \theta \left( (e,m) | \tau \right)^{-1} \ ,
\eea
with the theta function
\be
\theta(z|\tau) = -iq^{1/8}y^{1/2} \prod_{i=1}^\infty (1-q^i)(1-yq^i)(1-y^{-1}q^{i-1}) \ ,
\ee
where $q=e^{2\pi i \tau}$ and $y=e^{2\pi i z}$. The extra factors $e^{\frac{(N^2-1)\pi i}{12\tau}}$ and $\prod_{e<0}e^{\frac{2\pi i(e,m)}{\tau}}$ are multiplied by hand to the standard instanton series for $Z_{(3)}$ being a modular form. Again the former factor is interpreted as the leading term in high temperature expression of the 6d index, while the latter factor appears to be ambiguity which is not fixed in the 5d partition function.

The partition function in the presence of the defect is given by
\bea
    Z_{[1^N]} & =  \int [da] \prod_{e>0}2\sin \pi b^{-1} (e,a) \sum_{\sigma\in \mathcal{W}}\epsilon(\sigma)e^{-\frac{2\pi i}{\tau}b (\sigma(m),a)} e^{\frac{i\pi}{\tau}(a,a)} \times Z_{(3)} \\
    &= e^{-\frac{i\pi}{\tau} (m,m)( N + b^2 ) }  q^{-\frac{1}{24} (N^2-1) ( N/b^2 + 1 )} (q;q)_\infty \prod_{i,j=1}^N \frac{1}{(qz_i/z_j;q)_\infty} \\
    & = e^{-\frac{i\pi k(m,m)}{\tau}} q^{-c/24} \, \PE{\frac{q}{1-q}\left( \sum_{i,j=1}^N\frac{z_i}{z_j} - 1\right) } \ ,
\eea
where we write $z_j = e^{2\pi i m_j}$. The partition function can be recognized as the character of the vacuum module of the affine algebra $\widehat{\mathfrak{su}}(N)$ at level $k=-N-b^2$ and with associated Sugawara central charge $c = (N/b^2+1)(N^2-1)$. More precisely we have
\be
Z_{[1^N]} = e^{-\frac{i\pi k}{\tau} (m,m)}  \ \mathrm{Tr}_{V_0} \left(  q^{L_0-\frac{c}{24} }  \prod_j z_j^{h_j} \right) \ ,
\ee
where the vacuum module $V_0$ has highest affine weight $\hat\lb= k \hat\omega_0$. A summary of our conventions and a derivation of this result can be found in appendix~(\ref{asubsec:aff}). Note that in this case, we reproduce the vacuum character up to a small prefactor $e^{-\frac{i\pi k}{\tau}(m,m)}$ depending only on the level.


\subsection{Highest Weight Modules}\label{subsec:ahigh}

Let us now introduce a supersymmetric Wilson loop wrapping $S^1_{(1)}$. It is important that the support of this supersymmetric Wilson loop does not intersect the support $S^3_{(1)}$ of the surface defect. This means that the gauge symmetry is unbroken near the fixed point $(1)$ and we can introduce Wilson loops that is still in irreducible representations of $SU(N)$ labelled by the highest weight $\lambda$. From the perspective of the 6d superconformal index this corresponds to a codimension 4 defect orthogonal to the pre-existing codimension 4 defect - see figure~\ref{fig:summary}(c).

The presence of the supersymmetric Wilson loop changes only the contribution localized at $S^1_{(1)}$ to
\be
    Z_{(1)} = \prod_{e>0}2\sin \pi b^{-1}(e,a) \, \tr_{\lambda} (e^{2\pi i a/b} )
\ee
and thus the full partition function becomes
\bea
Z_{[1^N],\lb} = & Z_{[1^N]}  \times q^{ \frac{1}{2b^2}\left( (\rho,\rho) - (\rho+\lambda,\rho+\lambda) \right) }\, \chi_\lambda(e^{-2\pi i m}) \\
& = e^{-\frac{i\pi k}{\tau} (m,m)}q^{\Delta_\lb-c/24} \, \chi_{\lb}(z)\, \PE{\frac{q}{1-q}\left( \sum_{i,j=1}^N\frac{z_i}{z_j} - 1\right) } \ ,
\eea
where now the dimension is $\Delta_\lb = \frac{(\lb,\lb+2\rho)}{2(k+N)}$ and $\chi_\lb(z)$ denotes the character of the finite dimensional representation of $A_{N-1}$ with highest weight $\lb$. This is exactly the character of an irreducible module of the affine lie algebra at level $k = -N-b^2$ with affine highest weight $\hat \lb = k \hat\omega_0 + \lb$. As before, we refer the reader to appendix~(\ref{asubsec:aff}) for a summary of this result.


\section{Discussion}\label{sec:disc}

Let us conclude with a number of speculations and promising directions for further research.

Firstly, we believe that our computations provide an important step towards deriving the connection between the superconformal index of 4d $\cN=2$ theories of class $\mathcal{S}$ and topological quantum field theory~\cite{Gadde:2009kb,Gadde:2011ik}. Let us recall from~\cite{Gadde:2011ik} that in the Schur limit the superconformal index of the class $\cS$ theory corresponding to a Riemann surface with $n$ maximal punctures of type $\rho=[1^N]$ and genus $g$ can be expressed as
\be
\sum_{\lb} C_{\lb}^{2-2g-n} \psi_\lb(a_1) \ldots \psi_\lb(a_n)
\ee
where the summation is over the finite dimensional irreducible representations of $\mathfrak{g}$ and the expression for the structure constants $C_\lb$ and wavefunctions $\psi_\lb(a)$ can be found in~\cite{Gadde:2011ik}. Promoting the 4d superconformal index to the $S^1 \times S^3$ partition function
we observe in the case $\mathfrak{g}=A_{N-1}$ that
\begin{enumerate}
\item $C_\lb$ is the superconformal index of the $(2,0)$ theory in the presence of a codimension 4 defect labelled by dominant integral weight $\lambda$ orthogonal to the chiral algebra plane.
\item $\psi_\lb(a)$ is the superconformal index of the $(2,0)$ theory in the presence of a maximal codimension 2 defect wrapping the chiral algebra plane and an orthogonal codimension 4 defect labelled by $\lb$.   
\end{enumerate} 
Presumably this can be extended to non-maximal punctures of generic type $\rho$.

The above picture also suggests a concrete proposal for how to compute the wavefunctions $\psi_{\rho,\lb}(a_i,p,q,t)$ appearing in the superconformal index of class $\cS$ theories with general fugacities turned on. They should correspond to the superconformal index on $S^1 \times S^5$ with a codimension 2 defect of type $\rho$ wrapping say $S^1 \times S^3_{(1)}$ and codimension 4 defect of type $\lb$ wrapping $S^1\times S^1_{(1)}$. Note that the two defects coincide only along $S^1$: $S^3_{(1)}$ and $S^1_{(1)}$ are Hopf-linked inside $S^5$. The 4d parameters are identified with the parameters of the 6d superconformal index as
\be\label{eq:4d-6d-parameters}
\{a_i,p,q,t\}_{4d} = \{z_i,q_2,q_3,(q_1q_2q_3)^{1/2}p^{-1} \}_{6d}
\ee
Some initial checks of this proposal are performed in~\cite{Bullimore:2014awa}. This observation is of mathematical interest as these wavefunctions should provide a complete set of eigenfunctions of the elliptic Ruijsenaars-Schneider integral system for the codimension 2 defect of maximal type $\rho$ and more generally its degenerations. 

It is natural to identify the wavefunction $\psi_\lb$ with the contribution to the 4d superconformal index of class $\cS$ theories from a disk with puncture. This can be understood by reformulating the 6d superconformal index in terms of the 4d superconformal index on $S^1 \times S^3_{(1)}$ together with a topologically twist along the two transverse directions involving $S^1_{(1)}$, which are identified with a disk with puncture.
The puncture corresponds to the insertion of the codimension 2 defect of type $\rho$. The boundary condition along the $S^1_{(1)}$ may be specified by the codimension 4 defect of type $\lambda$. 

Note that there is an additional parameter in the 6d superconformal index conjugate to $h_1+R_2$, which is turned off in the identification (\ref{eq:4d-6d-parameters}). This reflects the topological twisting along the Riemann surface. One can also check the chiral algebra limit $p\rightarrow (q_1q_2/q_3)^{1/2}$ of the 6d superconformal index corresponds precisely to the Schur limit $t\rightarrow q$ of the 4d $\cN=2$ superconformal index of the theory on the $S^1 \times S^3_{(1)}$.
This suggests that it is possible to enumerate the states contributing to the superconformal index of 4d theories of class $\cS$ and identify their six-dimensional origin. We hope to return to this question in future work.

For codimension 2 defects wrapping $S^1 \times S^3_{(1)}$, we could express the partition function manifestly as a 6d superconformal index only in the case of a maximal puncture $\rho=[1^N]$. For more generic punctures, although we could find an integral expression and perform the integral explicitly, we could not transform the instanton contribution $Z_{(3)}^\rho$ from the third fixed point from a weak-coupling expansion in $\tilde{q}=e^{-2\pi i /\tau}$ to an expansion in the 6d fugacity $q = e^{2\pi i \tau}$. It will be interesting to find the modular property of $Z_{(3)}^\rho$ for a generic $\rho$ and compute the superconformal index.
Based on previous work, we would expect to find the characters of modules of $W(\rho)$ - algebras, which are obtained from the affine algebra $\widehat{\mathfrak{su}}(N)$ by Drinfeld-Sokolov reduction~\cite{Beem:2014rza}. As shown in \cite{Gaiotto:2012xa}, one can also reduce the flavor symmetry $\mathfrak{su}(N)$ of the maximal defect $\rho = [1^N]$ by coupling to a 4d $\cN=2$ linear quiver tail by gauging the global $\mathfrak{su}(N)$ symmetry and then Higgsing the theory by giving vacuum expectation values to the bifundamental hypermultiplets. 

Finally, a complementary approach to computing the 6d superconformal index is to use the 5d gauge theory on $S^1\times \mathbb{CP}^2$ by reducing the 6d (2,0) theory on $S^1\times S^5$ along the Hopf fiber of $S^5$. In this case, the partition function including the non-perturbative instanton contributions is expressed manifestly in the form of a 6d index, without the need for performing a modular transformation on $\tau$. The partition function with codimension 2 defects of generic type $\rho$ could be computed by this method and compared to the characters of $W(\rho)$ - algebra. This could also allow the 6d superconformal index with defects to be computed in the case of general fugacities.


\section*{Acknowledgements}

It is a pleasure to thank Nikolay Bobev, Davide Gaiotto, Jaume Gomis and Peter Koroteev for useful discussions. MB gratefully acknowledges support from the Perimeter Institute for Theoretical Physics and IAS Princeton through the Martin A. and Helen Choolijan Membership. HC gratefully acknowledges support from the Perimeter Institute for Theoretical Physics, the organizers of ``Exact Results in SUSY Gauge Theories in Various Dimensions'' at CERN and also CERN-Korea Theory Collaboration funded by National Research Foundation (Korea) for the hospitality and support. Research at the Perimeter Institute is supported by the Government of Canada and by the Province of Ontario through the Ministry of Research and Innovation.

\appendix


\section{Chiral Algebras}\label{asec:chiral}


\subsection{Conventons}\label{asubsec:conv}

Let us summarize our conventions for $\mathfrak{g} = A_{N-1}$. We choose the standard metric $(\, , \,)$ on $\mathfrak{h}^*$ normalized such that the length $(e,e)=2$ for all roots $e$ and identify $\mathfrak{h} = \mathfrak{h^*}$. The simple roots denoted $e_j$ are dual to the fundamental weights $\omega_j$ i.e.  $(e_i,\omega_j)=\delta_{ij}$. The Weyl vector is the sum of the fundamental weights $\rho = \sum_{j=1}^{N-1} \omega_j$ and has norm
\be
(\rho,\rho) = \frac{1}{12}N(N^2-1)\,.
\ee
The weights of the fundamental representation are
\bea
h_j = \omega_1 - e_1 - \ldots - e_{j-1} \qquad j = 1,\ldots,N\, ,
\eea
and obey $(h_i,h_j)=\delta_{ij}-1/N$ and $\sum_{j=1}^N h_j=0$. The weights of the rank-$r$ skew  tensor representation are then given by $ h_{j_1}+\ldots+h_{j_r}$ for $1\leq j_1<\ldots<j_r \leq N$. The positive roots are $h_i-h_j$ for $i<j$ and the simple roots are $e_j=h_j-h_{j+1}$. The Weyl group $S_N$ acts by permutations of $h_1,\ldots,h_N$.

We will often represent elements of $\mathfrak{g}$ by traceless anti-hermitian matrices
\be
h_j = i \, \mathrm{diag} \Big( -\frac{1}{N}, \ldots, \underbrace{1-\frac{1}{N}}_j,\ldots,-\frac{1}{N} \, \Big) \,
\ee
with $(a,b)=-\tr(ab)$ on $\mathfrak{h}$. For an element $a \in \mathfrak{h}$ we define its components by $a_j = (a,h_j)$ so that $\sum_{j=1}^N a_j = 0$ and $(a,a)=\tr(a^2)=a_1^1+ \ldots + a_N^2$.


\subsection{$W_N$ - algebra Characters}\label{asubsec:wn}

In this appendix, we will summarize the spectrum of simple modules of the $W$-algebra of type $\mathfrak{g} = A_{N-1}$, which we have called the $W_N$ - algebra in the main text, following closely reference~\cite{Drukker:2010jp}.

The $W_N$-algebra is generated by holomorphic currents $W_j(z)$ of spin $j=2,\ldots,N$.  The holomorphic current $W_2(x) = T(x)$ is identified with the stress energy tensor and generates a Virasoro subalgebra with central charge $c$ that can be parametrized by
\bea
c & = (N-1) + 12 (Q,Q) \\
& = (N-1)+(b+b^{-1})^2N(N^2-1)
\eea
where $Q = (b+b^{-1})\rho$ and $b>0$.

The simple modules $V_\mu$ are highest weight modules labelled by an element $\mu \in \mathfrak{h}$ called the momentum. They are constructed from a Verma module with chiral primary of dimension
\be
\Delta(\al) = (Q,\mu)-\frac{1}{2}(\mu,\mu)
\ee
by subtracting the descendants of any null vectors. The simple modules are sometimes classified crudely as non-degenerate, semi-degenerate or fully degenerate, depending the structure of null vectors appearing in the Verma module.

To construct simple modules we first choose a homomorphism $\rho:\mathfrak{su}(2) \to \mathfrak{g}$. This can be labelled by a partition $[n_1,\ldots,n_s]$ with $\sum_{j=1}^sn_j=N$ and by convention $n_i \leq n_j$ if $i<j$. This specifies how the fundamental representation of $\mathfrak{g}$ decomposes $N \to n_1 + \cdots + n_s$ under the image of the homomorphism $\mathrm{Im}(\rho) \subset \mathfrak{g}$. The stabilizer of $\mathrm{Im}(\rho)$ in $\mathfrak{g}$ is the subalgebra
\be
\mathfrak{l} = \mathfrak{s}( \mathfrak{u}(n_1) \oplus \cdots \oplus \mathfrak{u}(n_s))
\ee
with
\be
\Delta_\mathfrak{l} = \bigcup_{j=1}^s \Delta_j
\ee
where $\Delta_j$ is generated by the subset of simple roots $\{e_{r_j},\ldots,e_{r_{j+1}-1}\}$ with $r_j = n_1+\cdots+n_j$. The corresponding spaces of positive roots are denoted by $\Delta^+_\mathfrak{l}$ and $\Delta_j^+$ with Weyl vectors $\rho_\mathfrak{l}$ and $\rho_{n_j}$ defined as half the sum of the positive roots therein.

Given a homomorphism $\rho : \mathfrak{su}(2) \to \mathfrak{g}$, a simple module is constructed by starting from a Verma module with momentum
\be
\mu = Q + m - (\rho_{\mathfrak{l}} + \lb_1)/b -b (\rho_{\mathfrak{l}} + \lb_2)
\ee
where $m$ is an imaginary element of $\mathfrak{h}$ obeying $(m,\rho_\mathfrak{l})=0$ and $\lb_1$ and $\lb_2$ are dominant integral weights of $\mathfrak{l} \subset \mathfrak{g}$. The latter obey the conditions $(\lb_1,e) \geq 0$ for all $e \in \Delta_\mathfrak{l}^+$. There is in general an intricate structure of overlapping Verma modules generated by the null vectors. The character of the simple module obtained by subtracting the descendants of the null vectors is
\be
\tr_{V_\mu} \left( q^{L_0 - c/24} \right)  = \frac{q^{\Delta(\mu)-\frac{1}{2}(Q,Q)}}{\eta(\tau)^{N-1}} \sum_{w \in \cW_\mathfrak{l}} \ep(w) q^{-\left( w(\rho_\mathfrak{l}+\lb_1) - (\rho_\mathfrak{l} + \lb_1), \rho_\mathfrak{l}+\lb_2 \right)}  \,.
\ee
where $\cW_\mathfrak{l}$ is the Weyl group of $\mathfrak{l}$. The term in this formula with $w$ the identity element is the contribution from the full Verma module. The terms where $w$ is a reflection by a simple root in $\Delta_{\mathfrak{l}}$ have $\ep(w)=-1$ and subtract Verma modules generated by null vectors. The remaining terms account for  intersections of Verma modules and are fixed by invariance under $\cW_\mathfrak{l}$.

Let us now consider some extreme examples. Firstly, we consider the partition $\rho = [1^N]$ so that $\mathfrak{l} = \mathfrak{s}(\mathfrak{u}(1) \oplus \cdots \oplus \mathfrak{u}(1))$ and hence $\rho_{\mathfrak{l}}=0$. The parameter $m$ is now any imaginary element of $\mathfrak{h}$ and setting $\lb_1 = \lb_2=0$ we obtain
\be
\tr_{V_\mu} \left( q^{L_0 - c/24} \right)  = \frac{q^{-\frac{1}{2}(m,m)}}{\eta(\tau)^{N-1}} \,.
\ee
In this case, there are no null states and we retain the full Verma module. For this reason, these modules are called non-degenerate. At the other extreme we can consider the partition $\rho=[N]$ so that $\mathfrak{l} =\mathfrak{g}$. In this cases we must have $m=0$ and $\lb_1$ and $\lb_2$ are dominant integral weights of $\mathfrak{g}$. The character is
\be
\tr_{V_\mu} \left( q^{L_0 - c/24} \right)  = \frac{q^{\Delta(\mu)-\frac{1}{2}(Q,Q)}}{\eta(\tau)^{N-1}} \sum_{w \in \cW} \ep(w) q^{-\left( w(\rho+\lb_1) - (\rho + \lb_1), \rho+\lb_2 \right)}  \,.
\ee
where we have the momentum $\mu=-\lb_1/b-b\lb_2$. In this case, we have the maximum number of null vectors and the simple modules are called fully degenerate. In particular, the vacuum module corresponds to the case $\lb_1=\lb_2=0$. All other simple modules are broadly referred to as semi-degenerate.


\subsection{Affine Characters}\label{asubsec:aff}

We now consider some simple modules of the affine algebra $\widehat{\mathfrak{g}}$ with level in the regime $k = -N - \epsilon$ with $\epsilon >0$. This is generated by spin-1 holomorphic currents $J^a(x)$ with $a = 1,\ldots,N^2-1$. The Sugawara construction provides a Virasoro subalgebra with central charge
\be
c =\frac{(N^2-1)k}{k+N} = (N^2-1)(N/\ep-1) \, .
\ee

We will consider simple highest weight modules $V_{\hat\lb}$ labelled by a highest affine weight $\hat \lb$. The components of an affine weight are denoted by $\hat \lb = (\lb , k , n)$ where $\lb \in \mathfrak{h}$ is a finite weight, $k$ is the level and $n$ is the component dual to the generator $-L_0$. There is a metric denoted by $(\hat \lb ,\hat \lb') = (\lb,\lb') + k n'+ n k'$. We use the common abuse of notation and write $ \lb = (\lb , 0,0)$. It is convenient to introduce the fundamental affine weights which have components $\hat\omega_0 = (0,1,0)$ and $\hat\omega_j  = (\omega_j,1,0)$ for $j=1,\ldots,N-1$. An affine weight that is a linear combination of the fundamental affine weights can be written
\be
\hat\lb = \sum_{j=0}^{N-1}\lb_j\hat\omega_j  = k \hat\omega_0 + \lb
\ee
where $k = \sum_{j=0}^{N-1} \lb_j$. In particular, we introduce the affine Weyl vector $\hat \rho = \sum_{j=0}^{N-1} \hat \omega_j = N \hat\omega_0 + \rho$ and in components $\hat\rho = (\rho,N,0)$.

Introducing $\delta = (0,0,1)$, the affine roots are all of the form $\hat e = e + n \delta$ where $e \in \Delta$ is a finite root $\mathfrak{g}$. We will need the sets of positive and positive real affine roots, which are defined as
\bea
\hat\Delta^+ & =   \{ \al \in \Delta^+\} \cup \{ \al+n \delta \, | \, \al \in \Delta, n >0  \} \cup \{ n\delta \, | \, n>0\}   \\
\hat\Delta_{(\mathrm{re})}^+ & = \{ \al \in \Delta^+\} \cup \{ \al+n \delta \, | \, \al \in \Delta, n >0  \} \, .
\eea
The elements of $\hat\Delta_{(\mathrm{re})}^+$ have multiplicity one, whereas the imaginary roots $\{ n\delta \, | \, n>0\}$ have multiplicity $N-1$.

The characters of simple modules with highest affine weight $\hat\lb$ can be computed using the Kazhdan-Lusztig formula provided $ k \neq -N$. Given an affine weight $\hat\lb$ we define the following subset of the positive real affine roots
\be
\hat\Delta_{(\mathrm{re})}^+ (\hat\lb) = \{ \, \hat e \in \hat\Delta_{(\mathrm{re})}^+ \, | \, ( \hat \rho + \hat \lb , \hat e ) \in \bZ \, \} \, .
\ee
These roots generates a Coxeter group with associated Kazdan-Lusztig polynomials. In  the case that $k = -N-\epsilon$ with $\epsilon>0$ it has two components
\bea
& \quad (\rho + \lb , \al) \in \bZ  \quad && \mathrm{for} \quad \al \in \Delta^+ \\
& \quad (\rho+\lb , \al) - n \epsilon \in \bZ \quad  && \mathrm{for} \quad \al \in \Delta, \, n >0\, .
\eea
Here we want to consider generic $\ep>0$ so that the second component can only be non-empty by tuning $\lb$ in a way that depends on $\ep$. We will not consider this scenario. Instead, we take $\lb$ to be a dominant integral weight of $\mathfrak{g}$. The first component then consists of all $e \in \Delta^+$ and the Coxeter group is simply the Weyl group $\cW$ of $\mathfrak{g}$. The Kazhdan-Lusztig formula for the formal character is
\bea
\mathrm{Ch}(V_{\hat\lb}) & = \sum_{w \in \cW} \ep(w)  \frac{e^{w(\lb+\rho)-\rho}}{ \prod\limits_{\hat\al \in \hat\Delta_+} (1-e^{-\hat\al})^{\mathrm{mult}(\hat\al) } } \\
& = \frac{\sum_{w \in W}e^{w(\rho+\lb) - \rho}}{\prod\limits_{n > 0}(1-e^{-n\delta})^{N-1} \prod\limits_{\al \in \Delta^+}(1-e^{-\al}) \prod\limits_{\substack{\al \in \Delta \\ n \geq 1 }}(1-e^{-\al-\delta})  } \\
& = \frac{\chi_{\lb}(e^{-\al} )}{\prod\limits_{n > 0}(1-e^{-n\delta})^{N-1}  \prod_{\substack{\al \in \Delta \\ n \geq 1 }}(1-e^{-\al-\delta})  } \\
\eea
where we have assumed that only non-affine weight $\lb$ appears in the numerator. In passing to the third line we have used the Weyl denominator formula, and $\chi_{\lb}$ denotes the character of the simple finite dimensional module of $\mathfrak{g}$ with highest weight $\lb$. In particular, the vacuum module of $\widehat{\mathfrak{g}}$ with level $k$ corresponds to $\lb=0$ and $\hat \lb = k \,  \hat\omega_0$.

To compute the physical character we replace the formal expression $e^{-\al-n\delta}$ where $\al = \sum_j \ell_j h_j$ by the monomial $q^n \prod_j \mu_j^{\ell_j}$. Recall that $h_j$ are the weights of the fundamental representation and so $\ell_j$ components of $\al$ in the orthogonal basis. Therefore, we have
\bea
\mathrm{Tr}_{V_{\hat \lb}}\Bigg(q^{L_0 - \frac{c}{24}}\prod_{j=1}^N \mu_j^{h_j}\Bigg) & = \frac{q^{-\frac{c}{24}} \chi_{\lb}(\mu)  }{ \prod\limits_{n > 0}(1-q^n)^{N-1} \prod\limits_{\substack{i \neq j  \\ n \geq 0} } (1-q^n \mu_i / \mu_j) } \\
& = q^{- \frac{c}{24}} \frac{(q,q) \chi_{\lb}(\mu) }{\prod_{i,j=1}^N (q\mu_i/\mu_j,q)} \, .
\eea


\section{$S^5$ partition function and codimension 2 defects}\label{sec-S5-partitionftn}
One can construct the 5d maximal SYM theory on (squashed) $S^5$ from
the 6d (2,0) theory on $S^5 \times S^1$ by dimensional reduction along
the $S^1$. 
We first reduce the abelian (2,0) theory to five-dimensions and find
its non-abelian generalization. 
Let us consider the 6d theory defined on the curved metric
\be
	ds_6^2 = e^{-\frac{2}{3}\Phi}ds_5^2 + e^{\frac{4}{3}\Phi} (dt + e^{-\Phi}C)^2
\ee
where $t$ is the Euclidean time and
\be\label{eq-squashed-metric}
	e^{\frac{4}{3}\Phi} = 1-n_i^2a_i^2 \,, \quad C = \frac{in_i^2a_id\phi_i}{1-n_i^2a_i^2} \,, \quad
	ds_5^2 = (1-n_i^2a_i^2)^{\frac{1}{2}} \left[ dn_i^2 + n_i^2d\phi_i^2 + \frac{(a_in_i^2d\phi_i)^2}{1-n_i^2a_i^2} \right]
\ee
Here $ w_i = 1 + a_i$ are chemical potentials for the $U(1)^3$ rotation of the holomorphic coordinates $z_i= n_i e^{i\phi},$ $(n_1^2+n_2^2+n_3^2 = 1)$.
The dimensional reduction along the time circle gives rise to the 5d theory on the squashed five-sphere
with squashing parameters $\omega_j$.
The background 'dilaton' $\Phi$ and 'RR gauge field' $C_\mu$ are also turned on. 

We restrict the 5d reduction such that it preserves two supercharges $Q$ and $Q^\dagger$ used to define the 6d superconformal index.
The 5d gauge theory action is uniquely determined under this reduction.
The explicit action can be found in \cite{Kim:2012qf}.
The 5d supercharge $Q$ satisfies the following algebra:
\be
	\{Q,Q^\dagger \} \sim -\frac{3(R_1+R_2)}{2} - \mu (R_1-R_2) - \sum_{i=1}^3 \omega_ih_i \ .
\ee

The partition function of the 5d theory can be computed using localization.
The saddle points of the path integral are given by a constant scalar vev $\langle \phi \rangle=a$ in the vectormultiplet and
the singular instantons at the fixed points of the Killing vector $\xi \equiv  \sum_{i=1}^3 \omega_ih_i $.
The final result is \cite{Kim:2012qf}
\be
	Z = \frac{1}{N!}\int [da]e^{\frac{2\pi^2}{\beta \omega_1\omega_2\omega_3}(a,a)} \prod_{i=1}^3 Z_{\rm pert}^{(i)} Z_{\rm inst}^{(i)} \ ,
\ee
where $i$'s label three fixed circles $S^1_{(i)} \subset S^5$.
The 1-loop contribution is factorized into three fixed circle contributions $Z^{(i)}_{\rm pert}$ and,
by collecting all of them, one obtains \cite{Lockhart:2012vp,Imamura:2012xg,Kim:2012qf}
\bea\label{eq-perturbative-determinant}
	\hspace{-0.5cm}\prod_{i=1}^3Z_{\rm pert}^{(i)}  &=
	\prod_{e\in \Delta} \prod_{p,q,r=0}^\infty
	\frac{( p\omega_1+q\omega_2+r\omega_3+(e,a) )' ( (p+1)\omega_1+(q+1)\omega_2+(r+1)\omega_3+(e,a) )}
	{(p \omega_1+q\omega_2+r\omega_3+\tilde{\mu}+(e,a) )
	( (p+1)\omega_1+(q+1)\omega_2+(r+1)\omega_3-\tilde{\mu}+(e,a) )}  \\
	&= \left( \frac{\lim_{x\rightarrow 0}S_3(x)/x}{S_3(\tilde{\mu})} \right)^N \prod_{e>0} 
	\frac{S_3\big((e,a)|\vec{\omega}\big)S_3\big(-(e,a)|\vec{\omega}\big)}
	{S_3\big(\tilde{\mu}+(e,a)|\vec{\omega}\big)S_3\big(\tilde{\mu} - (e,a)|\vec{\omega}\big)} \ ,
\eea
where $\tilde{\mu} \equiv \mu + \frac{\omega_1+\omega_2+\omega_3}{2}$
and $S_3(x|\vec{\omega})$ is the triple-Sine function with $\vec{\omega}=(\omega_1,\omega_2,\omega_3)$.
The prime in the first line denotes that the modes with $p=q=r=e=0$ are removed.
The instanton contribution $Z_{\rm inst}^{(i)}$ at each fixed point coincides with the 5d Nekrasov instanton partition function on $S^1\times \mathbb{C}^2$ with Omega deformation.

Let us now turn to the codimension 2 defects on $S^5$, which are related to the codimension 2 operators in the 6d (2,0) theory. The BPS defects can be supported on $S^3 \subset S^5$. For simplicity, let us stick to the maximal SYM theory with $U(N)$ (or $SU(N)$) gauge group on round $S^5$. These defects are defined by specifying a singular behavior of the gauge field as one approaches their location.
Near the defects, we parametrize two normal directions by a complex coordinate $z=re^{i\theta}$ where $\theta$ is one of the angle coordinates in $S^5$.
Then the defect is defined with a gauge field which behaves around the defect as
\be
	A_\mu dx^\mu
	\sim  \vec{m} d\theta \equiv  {\rm diag}(m_1,m_2,\cdots, m_N) d\theta \ .
\ee
Here $\vec{m}$ is a collection of monodromy parameters. For $SU(N)$, it obeys $\sum_i m_i = 0$.
The corresponding field strength takes the form
\be\label{eq-codim-two-flux}
	F = \vec{m}\, \frac{\delta(r)}{r}  * d\Omega_{S^3} \ ,
\ee
where $* d\Omega_{S^3}$ is the Hodge dual of the three-sphere volume form.

Let us now derive the classical action in the presence of the defect.
We use the off-shell supersymmetry formulation of the 5d SYM studied in \cite{Hosomichi:2012ek,Kim:2012ava}.
We focus on the round $S^5$ background.
It turns out that the codimension 2 defects preserve the supercharge $Q$ used in the localization.
The BPS condition from the gaugino variation is given by
\be
	\frac{1}{2}F_{\mu\nu}\gamma^{\mu\nu}\epsilon -i D_\mu \phi^\mu \epsilon
	+ \phi \sigma^3\epsilon + i D^I\sigma^I \epsilon = 0 \ .
\ee
The Killing spinor $\epsilon$ for the supercharge $Q$ satisfies the following conditions
\be
	\epsilon^\dagger \gamma^\mu \epsilon = v^\mu \,, \quad \epsilon^\dagger\gamma^{\mu\nu}\epsilon = iJ^{\mu\nu}
	\,, \quad \sigma^3\epsilon = \epsilon \ ,
\ee
where $v^\mu$ is the Killing vector along the Hopf fiber of $S^5$ and $J^{\mu\nu}$ is the K\"ahler form of
$\mathbb{CP}^2$ base.
The solution to the BPS equation on the background flux (\ref{eq-codim-two-flux}) is given by
\be
	F = \vec{m} \, \frac{\delta(r)}{r}  * d\Omega_{S^3} \,, \quad \phi = a \,, \quad D^3 = \vec{m} \, \frac{\delta(r)}{r} + i a
	\,, \quad D^{I=1,2} = 0 \ , 
\ee
where $a$ is a constant Hermitian matrix taking values in the Lie algebra of the gauge group.
Plugging this solution into the the action, one obtains the classical action with the codimension 2 defect
\be
	e^{-S_0}\,, \quad 
	S_0 = \frac{1}{g^2_{YM}} \int d^5x \sqrt{g} \ {\rm Tr} \! \left[\frac{1}{4} F_{\mu\nu}F^{\mu\nu} 
	-\frac{1}{2} D^I D^I - i D^3\phi +\frac{5}{2} \phi^2 \right]
	= \frac{2\pi^2}{\beta} (a-2i\vec{m},a) \ .
\ee
In the main context, we would use the convention $a \rightarrow ia$ by analytic continuation.

We would not perform an explicit localization computation in the presence of codimension 2 defect. However, turning on the squashing parameters, we expect that the path integral again localizes to three fixed points and the full partition function takes the form of products of three fixed point contributions.
The contributions at the fixed points can be computed using the results on the local $S^1\times \mathbb{C}^2$, which are explained in sections~\ref{subsec:comp}$\,\sim$~\ref{subsec:codim2}.

\bibliographystyle{JHEP}
\bibliography{6dindex}

\providecommand{\href}[2]{#2}\begingroup\raggedright\begin{thebibliography}{10}

\bibitem{Gaiotto:2009we}
D.~Gaiotto, {\it {N=2 dualities}},  {\em JHEP} {\bf 1208} (2012) 034,
  [\href{http://arxiv.org/abs/0904.2715}{{\tt arXiv:0904.2715}}].

\bibitem{Gaiotto:2009hg}
D.~Gaiotto, G.~W. Moore, and A.~Neitzke, {\it {Wall-crossing, Hitchin Systems,
  and the WKB Approximation}},  \href{http://arxiv.org/abs/0907.3987}{{\tt
  arXiv:0907.3987}}.

\bibitem{Alday:2009aq}
L.~F. Alday, D.~Gaiotto, and Y.~Tachikawa, {\it {Liouville Correlation
  Functions from Four-dimensional Gauge Theories}},  {\em Lett.Math.Phys.} {\bf
  91} (2010) 167--197, [\href{http://arxiv.org/abs/0906.3219}{{\tt
  arXiv:0906.3219}}].

\bibitem{Gadde:2009kb}
A.~Gadde, E.~Pomoni, L.~Rastelli, and S.~S. Razamat, {\it {S-duality and 2d
  Topological QFT}},  {\em JHEP} {\bf 1003} (2010) 032,
  [\href{http://arxiv.org/abs/0910.2225}{{\tt arXiv:0910.2225}}].

\bibitem{Gadde:2011ik}
A.~Gadde, L.~Rastelli, S.~S. Razamat, and W.~Yan, {\it {The 4d Superconformal
  Index from q-deformed 2d Yang-Mills}},  {\em Phys.Rev.Lett.} {\bf 106} (2011)
  241602, [\href{http://arxiv.org/abs/1104.3850}{{\tt arXiv:1104.3850}}].

\bibitem{Dimofte:2011ju}
T.~Dimofte, D.~Gaiotto, and S.~Gukov, {\it {Gauge Theories Labelled by
  Three-Manifolds}},  {\em Commun.Math.Phys.} {\bf 325} (2014) 367--419,
  [\href{http://arxiv.org/abs/1108.4389}{{\tt arXiv:1108.4389}}].

\bibitem{Dimofte:2011py}
T.~Dimofte, D.~Gaiotto, and S.~Gukov, {\it {3-Manifolds and 3d Indices}},  {\em
  Adv.Theor.Math.Phys.} {\bf 17} (2013) 975--1076,
  [\href{http://arxiv.org/abs/1112.5179}{{\tt arXiv:1112.5179}}].

\bibitem{Gadde:2013sca}
A.~Gadde, S.~Gukov, and P.~Putrov, {\it {Fivebranes and 4-manifolds}},
  \href{http://arxiv.org/abs/1306.4320}{{\tt arXiv:1306.4320}}.

\bibitem{Kim:2012ava}
H.-C. Kim and S.~Kim, {\it {M5-branes from gauge theories on the 5-sphere}},
  {\em JHEP} {\bf 1305} (2013) 144, [\href{http://arxiv.org/abs/1206.6339}{{\tt
  arXiv:1206.6339}}].

\bibitem{Lockhart:2012vp}
G.~Lockhart and C.~Vafa, {\it {Superconformal Partition Functions and
  Non-perturbative Topological Strings}},
  \href{http://arxiv.org/abs/1210.5909}{{\tt arXiv:1210.5909}}.

\bibitem{Kim:2012qf}
H.-C. Kim, J.~Kim, and S.~Kim, {\it {Instantons on the 5-sphere and
  M5-branes}},  \href{http://arxiv.org/abs/1211.0144}{{\tt arXiv:1211.0144}}.

\bibitem{Kallen:2012cs}
J.~K{\"a}ll{\'e}n and M.~Zabzine, {\it {Twisted supersymmetric 5D Yang-Mills
  theory and contact geometry}},  {\em JHEP} {\bf 1205} (2012) 125,
  [\href{http://arxiv.org/abs/1202.1956}{{\tt arXiv:1202.1956}}].

\bibitem{Hosomichi:2012ek}
K.~Hosomichi, R.-K. Seong, and S.~Terashima, {\it {Supersymmetric Gauge
  Theories on the Five-Sphere}},  {\em Nucl.Phys.} {\bf B865} (2012) 376--396,
  [\href{http://arxiv.org/abs/1203.0371}{{\tt arXiv:1203.0371}}].

\bibitem{Kallen:2012va}
J.~K{\"a}ll{\'e}n, J.~Qiu, and M.~Zabzine, {\it {The perturbative partition
  function of supersymmetric 5D Yang-Mills theory with matter on the
  five-sphere}},  {\em JHEP} {\bf 1208} (2012) 157,
  [\href{http://arxiv.org/abs/1206.6008}{{\tt arXiv:1206.6008}}].

\bibitem{Imamura:2012xg}
Y.~Imamura, {\it {Supersymmetric theories on squashed five-sphere}},  {\em
  PTEP} {\bf 2013} (2013) 013B04, [\href{http://arxiv.org/abs/1209.0561}{{\tt
  arXiv:1209.0561}}].

\bibitem{Imamura:2012bm}
Y.~Imamura, {\it {Perturbative partition function for squashed $S^5$}},
  \href{http://arxiv.org/abs/1210.6308}{{\tt arXiv:1210.6308}}.

\bibitem{Douglas:2010iu}
M.~R. Douglas, {\it {On D=5 super Yang-Mills theory and (2,0) theory}},  {\em
  JHEP} {\bf 1102} (2011) 011, [\href{http://arxiv.org/abs/1012.2880}{{\tt
  arXiv:1012.2880}}].

\bibitem{Lambert:2010iw}
N.~Lambert, C.~Papageorgakis, and M.~Schmidt-Sommerfeld, {\it {M5-Branes,
  D4-Branes and Quantum 5D super-Yang-Mills}},  {\em JHEP} {\bf 1101} (2011)
  083, [\href{http://arxiv.org/abs/1012.2882}{{\tt arXiv:1012.2882}}].

\bibitem{Beem:2014kka}
C.~Beem, L.~Rastelli, and B.~C. van Rees, {\it {W Symmetry in six dimensions}},
   \href{http://arxiv.org/abs/1404.1079}{{\tt arXiv:1404.1079}}.

\bibitem{Drukker:2010jp}
N.~Drukker, D.~Gaiotto, and J.~Gomis, {\it {The Virtue of Defects in 4D Gauge
  Theories and 2D CFTs}},  {\em JHEP} {\bf 1106} (2011) 025,
  [\href{http://arxiv.org/abs/1003.1112}{{\tt arXiv:1003.1112}}].

\bibitem{Kinney:2005ej}
J.~Kinney, J.~M. Maldacena, S.~Minwalla, and S.~Raju, {\it {An Index for 4
  dimensional super conformal theories}},  {\em Commun.Math.Phys.} {\bf 275}
  (2007) 209--254, [\href{http://arxiv.org/abs/hep-th/0510251}{{\tt
  hep-th/0510251}}].

\bibitem{Bhattacharya:2008zy}
J.~Bhattacharya, S.~Bhattacharyya, S.~Minwalla, and S.~Raju, {\it {Indices for
  Superconformal Field Theories in 3,5 and 6 Dimensions}},  {\em JHEP} {\bf
  0802} (2008) 064, [\href{http://arxiv.org/abs/0801.1435}{{\tt
  arXiv:0801.1435}}].

\bibitem{Kim:2013nva}
H.-C. Kim, S.~Kim, S.-S. Kim, and K.~Lee, {\it {The general M5-brane
  superconformal index}},  \href{http://arxiv.org/abs/1307.7660}{{\tt
  arXiv:1307.7660}}.

\bibitem{Bobev:2015kza}
N.~Bobev, M.~Bullimore, and H.-C. Kim, {\it {Supersymmetric Casimir Energy and
  the Anomaly Polynomial}},  {\em JHEP} {\bf 09} (2015) 142,
  [\href{http://arxiv.org/abs/1507.08553}{{\tt arXiv:1507.08553}}].

\bibitem{Nekrasov:2002qd}
N.~A. Nekrasov, {\it {Seiberg-Witten prepotential from instanton counting}},
  {\em Adv.Theor.Math.Phys.} {\bf 7} (2004) 831--864,
  [\href{http://arxiv.org/abs/hep-th/0206161}{{\tt hep-th/0206161}}].

\bibitem{Nekrasov:2003rj}
N.~Nekrasov and A.~Okounkov, {\it {Seiberg-Witten theory and random
  partitions}},  \href{http://arxiv.org/abs/hep-th/0306238}{{\tt
  hep-th/0306238}}.

\bibitem{Losev:2003py}
A.~S. Losev, A.~Marshakov, and N.~A. Nekrasov, {\it {Small instantons, little
  strings and free fermions}},  \href{http://arxiv.org/abs/hep-th/0302191}{{\tt
  hep-th/0302191}}.

\bibitem{Shadchin:2004yx}
S.~Shadchin, {\it {Saddle point equations in Seiberg-Witten theory}},  {\em
  JHEP} {\bf 0410} (2004) 033, [\href{http://arxiv.org/abs/hep-th/0408066}{{\tt
  hep-th/0408066}}].

\bibitem{2003math......6164N}
A.~{Narukawa}, {\it {The modular properties and the integral representations of
  the multiple elliptic gamma functions}},  {\em ArXiv Mathematics e-prints}
  (June, 2003) [\href{http://arxiv.org/abs/math/0306164}{{\tt math/0306164}}].

\bibitem{Young:2011aa}
D.~Young, {\it {Wilson Loops in Five-Dimensional Super-Yang-Mills}},  {\em
  JHEP} {\bf 1202} (2012) 052, [\href{http://arxiv.org/abs/1112.3309}{{\tt
  arXiv:1112.3309}}].

\bibitem{Assel:2012nf}
B.~Assel, J.~Estes, and M.~Yamazaki, {\it {Wilson Loops in 5d N=1 SCFTs and
  AdS/CFT}},  {\em Annales Henri Poincare} {\bf 15} (2014) 589--632,
  [\href{http://arxiv.org/abs/1212.1202}{{\tt arXiv:1212.1202}}].

\bibitem{Kim:2011mv}
H.-C. Kim, S.~Kim, E.~Koh, K.~Lee, and S.~Lee, {\it {On instantons as
  Kaluza-Klein modes of M5-branes}},  {\em JHEP} {\bf 1112} (2011) 031,
  [\href{http://arxiv.org/abs/1110.2175}{{\tt arXiv:1110.2175}}].

\bibitem{Gukov:2006jk}
S.~Gukov and E.~Witten, {\it {Gauge Theory, Ramification, And The Geometric
  Langlands Program}},  \href{http://arxiv.org/abs/hep-th/0612073}{{\tt
  hep-th/0612073}}.

\bibitem{Gomis:2007fi}
J.~Gomis and S.~Matsuura, {\it {Bubbling surface operators and S-duality}},
  {\em JHEP} {\bf 0706} (2007) 025, [\href{http://arxiv.org/abs/0704.1657}{{\tt
  arXiv:0704.1657}}].

\bibitem{Alday:2009fs}
L.~F. Alday, D.~Gaiotto, S.~Gukov, Y.~Tachikawa, and H.~Verlinde, {\it {Loop
  and surface operators in N=2 gauge theory and Liouville modular geometry}},
  {\em JHEP} {\bf 1001} (2010) 113, [\href{http://arxiv.org/abs/0909.0945}{{\tt
  arXiv:0909.0945}}].

\bibitem{Alday:2010vg}
L.~F. Alday and Y.~Tachikawa, {\it {Affine SL(2) conformal blocks from 4d gauge
  theories}},  {\em Lett.Math.Phys.} {\bf 94} (2010) 87--114,
  [\href{http://arxiv.org/abs/1005.4469}{{\tt arXiv:1005.4469}}].

\bibitem{Chacaltana:2012zy}
O.~Chacaltana, J.~Distler, and Y.~Tachikawa, {\it {Nilpotent orbits and
  codimension-two defects of 6d N=(2,0) theories}},  {\em Int.J.Mod.Phys.} {\bf
  A28} (2013) 1340006, [\href{http://arxiv.org/abs/1203.2930}{{\tt
  arXiv:1203.2930}}].

\bibitem{Tachikawa:2011dz}
Y.~Tachikawa, {\it {On W-algebras and the symmetries of defects of 6d N=(2,0)
  theory}},  {\em JHEP} {\bf 1103} (2011) 043,
  [\href{http://arxiv.org/abs/1102.0076}{{\tt arXiv:1102.0076}}].

\bibitem{Kanno:2011fw}
H.~Kanno and Y.~Tachikawa, {\it {Instanton counting with a surface operator and
  the chain-saw quiver}},  {\em JHEP} {\bf 1106} (2011) 119,
  [\href{http://arxiv.org/abs/1105.0357}{{\tt arXiv:1105.0357}}].

\bibitem{Gadde:2011uv}
A.~Gadde, L.~Rastelli, S.~S. Razamat, and W.~Yan, {\it {Gauge Theories and
  Macdonald Polynomials}},  {\em Commun.Math.Phys.} {\bf 319} (2013) 147--193,
  [\href{http://arxiv.org/abs/1110.3740}{{\tt arXiv:1110.3740}}].

\bibitem{Gaiotto:2012xa}
D.~Gaiotto, L.~Rastelli, and S.~S. Razamat, {\it {Bootstrapping the
  superconformal index with surface defects}},  {\em JHEP} {\bf 1301} (2013)
  022, [\href{http://arxiv.org/abs/1207.3577}{{\tt arXiv:1207.3577}}].

\bibitem{Gaiotto:2008ak}
D.~Gaiotto and E.~Witten, {\it {S-Duality of Boundary Conditions In N=4 Super
  Yang-Mills Theory}},  {\em Adv.Theor.Math.Phys.} {\bf 13} (2009) 721,
  [\href{http://arxiv.org/abs/0807.3720}{{\tt arXiv:0807.3720}}].

\bibitem{2008arXiv0812.4656F}
B.~{Feigin}, M.~{Finkelberg}, A.~{Negut}, and L.~{Rybnikov}, {\it {Yangians and
  cohomology rings of Laumon spaces}},  {\em ArXiv e-prints} (Dec., 2008)
  [\href{http://arxiv.org/abs/0812.4656}{{\tt arXiv:0812.4656}}].

\bibitem{2010arXiv1009.0676F}
M.~{Finkelberg} and L.~{Rybnikov}, {\it {Quantization of Drinfeld Zastava in
  type A}},  {\em ArXiv e-prints} (Sept., 2010)
  [\href{http://arxiv.org/abs/1009.0676}{{\tt arXiv:1009.0676}}].

\bibitem{Bullimore:2014awa}
M.~Bullimore, H.-C. Kim, and P.~Koroteev, {\it {Defects and Quantum
  Seiberg-Witten Geometry}},  {\em JHEP} {\bf 05} (2015) 095,
  [\href{http://arxiv.org/abs/1412.6081}{{\tt arXiv:1412.6081}}].

\bibitem{Nawata:2014nca}
S.~Nawata, {\it {Givental J-functions, Quantum integrable systems, AGT relation
  with surface operator}},  \href{http://arxiv.org/abs/1408.4132}{{\tt
  arXiv:1408.4132}}.

\bibitem{Wyllard:2010vi}
N.~Wyllard, {\it {Instanton partition functions in N=2 SU(N) gauge theories
  with a general surface operator, and their W-algebra duals}},  {\em JHEP}
  {\bf 1102} (2011) 114, [\href{http://arxiv.org/abs/1012.1355}{{\tt
  arXiv:1012.1355}}].

\bibitem{Pasquetti:2011fj}
S.~Pasquetti, {\it {Factorisation of N = 2 Theories on the Squashed 3-Sphere}},
   {\em JHEP} {\bf 1204} (2012) 120,
  [\href{http://arxiv.org/abs/1111.6905}{{\tt arXiv:1111.6905}}].

\bibitem{BBK2014}
N.~Bobev, M.~Bullimore, and H.-C. Kim, {\it {in progress}}, .

\bibitem{Beem:2014rza}
C.~Beem, W.~Peelaers, L.~Rastelli, and B.~C. van Rees, {\it {Chiral algebras of
  class S}},  \href{http://arxiv.org/abs/1408.6522}{{\tt arXiv:1408.6522}}.

\end{thebibliography}\endgroup


\providecommand{\href}[2]{#2}\begingroup\raggedright\endgroup

\end{document}